\documentstyle[eqsecnum,pre,preprint,aps,epsfig,rotating]{revtex}

\tightenlines

\newcommand{\bq}{\begin{equation}}
\newcommand{\eq}{\end{equation}}
\newcommand{\bqa}{\begin{eqnarray}}
\newcommand{\eqa}{\end{eqnarray}}

\begin{document}

\draft
\preprint{PM/01-19, hep-ph/0104205, corrected version}
\vspace{2cm}

\title{High energy behaviour of $\gamma\gamma \to f\bar f$ processes in
SM and MSSM}

\author{J. Layssac and F.M. Renard }

\address{Physique
Math\'{e}matique et Th\'{e}orique, UMR 5825\\
Universit\'{e} Montpellier
II,  F-34095 Montpellier Cedex 5.
}
\vspace{2cm}
\maketitle
\vspace{2cm}
\begin{abstract}

We compute the leading logarithms electroweak contributions
to $\gamma\gamma \to f\bar f$ processes in SM and MSSM. Several
interesting properties are pointed out,
such as the importance of the angular dependent terms, of the
Yukawa terms, and especially of the $\tan^2\beta$ dependence in the SUSY
contributions. These properties are complementary to those
found in $e^+e^-\to f\bar f$ .  
These radiative correction
 effects should be largely observable at future
high energy $\gamma\gamma$ colliders. Polarized beams would
bring interesting checks of the structure of the one loop corrections.
We finally discuss the need for two-loop calculations and
resummation. 
\end{abstract}
\vspace{2cm}

\section{Introduction.}

The projects of high energy and high luminosity $e^+e^-$ colliders
\cite{LC,CLIC} have recently motivated the study of the high energy
behaviour of the electroweak corrections to several $e^+e^-$
annihilation processes. Explicit computations of the linear and
quadratic logarithmic contributions to various observables have shown
remarkable properties which should be largely observable 
at these future machines and 
should provide deep tests of the different sectors 
(gauge, matter, scalar) of the Standard
Model (SM) as well as of its 
Supersymmetric extensions, like
the Minimal Supersymmetric Standard Model (MSSM) \cite{alllog,Kuhn}.
In fact, since many years, it was known that, in certain circumstances,
large logarithmic terms, in particular quadratic logarithms
can appear\cite{Sudakov,ln2}. The general features 
of the asymptotic one loop electroweak corrections have been
studied, a classification of the linear and
quadratic logarithms have been established, some two loop effects have
been computed and the possibility of resumming certain classes of
contributions have been discussed \cite{KPS,CCC,tl,Melles,DenPo}.\\

On another hand, the possibility of realizing 
high energy and high luminosity
$\gamma\gamma$ collisions at $e^+e^-$ colliders 
through the laser backscattering procedure 
is actively considered \cite{laser,ggstudies}.
One already knows that electroweak radiative corrections
to $\gamma\gamma\to f\bar f$ processes, both in the SM\cite{dendit}
and in the MSSM\cite{Zhou} are sizeable enough to 
be observable owing to the large luminosities expected
at these machines which should allow to reach an accuracy better 
than the percent level.\\

The purpose of the present paper is to report on 
a study of the high energy
behaviour of the electroweak corrections to the process
$\gamma\gamma\to f\bar f$ in SM and in MSSM,
performed along the same lines as those taken for
the aforementioned studies of the
$e^+e^-\to f\bar f$ processes. We will show that the
$\gamma\gamma\to f\bar f$ processes offer an independent
way to check the general properties of the asymptotic logarithmic
terms originating from the various sectors of the
electroweak interactions, and we will give precise numerical 
illustrations in order to see how they can be experimentally tested.
A great similarity with the properties of the
$e^+e^-\to f\bar f$ processes will appear and will allow us to
conclude that $\gamma\gamma\to f\bar f$ processes can
equally well contribute to the tests of the SM at high energies
and to the search for its possible modifications or extensions.\\ 

The contents of the paper is the following. In Section II we present the
dynamical contents in SM and in MSSM and 
we proceed with the computation of the
complete one-loop weak contributions in the asymptotic regime.
QED and QCD corrections are left aside as they depend on the detection
conditions and are usually included in specific Monte-Carlo programs
\cite{dendit}.
After having checked that the set of self-energy, vertex and box
diagrams which are retained in the high energy limit is
gauge-independent and satisfies photon current conservation, we
systematically work in the $\xi=1$ gauge. We check the convergence
of the separate
contributions of the various sectors (neutral gauge, charged gauge,
Yukawa) of the Standard Model (SM) , as well as of the additional SUSY
terms (gaugino, higgsino, additional Higgs bosons). 
We keep the single and
the quadratic logarithmic contributions. We separate the angular
independent corrections from the angular dependent ones. 
All these contributions are specified 
for the helicity amplitudes of the
process $\gamma\gamma\to f\bar f$; they are explicitly
given in analytical form in Appendix A and B. 
From these expressions it is then easy to compute the
various parts of the fully polarized $\gamma\gamma$ cross section.
This is what we present in Section III. We then compute
the effects on the various $\gamma\gamma$ 
observables and
we present and discuss the results in the SM and MSSM
cases. With the expected luminosity of LC and CLIC these
various contributions should be experimentally observable.
We then discuss the physics implications of the results
as well as the domain of validity of the one-loop computation 
and the need
of a two loop computation or a resummation at very high energies.
This output is summarized in the concluding Section IV.

\section{Dynamical 1-loop contents of $\gamma\gamma\to 
\lowercase{f \bar f}$
at high energy}

We found convenient to express all the results in terms of helicity
amplitudes \cite{JW} $F_{\lambda,\lambda',\tau,\tau'}$, 
$\lambda,\lambda',\tau,\tau'$ being the helicities of the two photons
and of the fermion, antifermion, respectively;
it is then
easy to get the expressions of the observables 
in polarized photon-photon collisions.\\

The Born term consists in 2 diagrams with fermion
exchange in the $t$ and $u$ channels. 
It is $\gamma\gamma$ symmetric; its amplitude,
in the high energy limit, is written in
Appendix A. It only contributes to the
$|\Delta\lambda|=2$ helicity amplitudes.\\

At one-loop, the list of diagrams (to be symmetrized by interchanging
the two photons) which contribute to the logarithmic terms
in the high energy limit is given in Fig.1a-c  for the SM case. 
In the MSSM case, the additional SUSY diagrams 
can be found in Fig.2a-b . 
We have checked that these contributions are ($\xi-1$)-independent and
that current conservation ($l^{\mu}J_{\mu}=0$) holds separately
for each photon. In Fig.1a-c,2a-b we have not
drawn the external (photon, fermion) self-energy diagrams which do not
contribute to the logarithmic terms, although they 
must be taken into account in
order to get cancelation of the divergences generated by the
internal fermion self-energy and by the triangular diagrams; box
diagrams are convergent.\\

The explicit expressions of the helicity amplitudes
in the high energy limit
are given, separately for each sector 
of the electroweak corrections, in an analytical
form in Appendix A. They are obtained 
by deriving the complete expressions of the amplitudes
in terms of Passarino-Veltman
functions \cite{PV}, and retaining only the asymptotic 
(logarithmic) parts of these functions (see Appendix B). 
In a second step we only retain the terms which contain
linear ($\ln s$) and quadratic ($\ln^2s$) logarithms,that we call
"\underline{leading terms}", neglecting
terms like $\ln(t/s)$, ....etc, that we call 
"\underline{non leading terms}".
During this procedure we have checked that the divergences and 
the fermion mass singularities cancel.
We have also separated the coefficients of the leading logarithms
which are $\theta$-independent from those which are
$\theta$-dependent ($\theta$ is the c.m. scattering angle).
We now discuss in turn these various terms.\\

\newpage

{\bf Standard Model corrections}\\

\underline{$\gamma$ and $Z$ sectors}\\

A first set of corrections is given by the internal fermion
self-energy, triangle and box diagrams of Fig.1a containing one $Z$
boson. The corresponding helicity amplitudes are given 
in eq.(\ref{gZm}) and
(\ref{gZp}) (terms proportional to ${[g^Z_{Vf}-g^Z_{Af}(2\tau)]^2
/ 4s^2_Wc^2_W}$). One can check in eqs.(\ref{gZmlt}),(\ref{gZplt})
that the leading terms of the
$|\Delta\lambda|=2$ helicity amplitude combine in
an angular independent factor proportional 
to $[\ln^2(s/ M^2_Z)-3\ln (s/ M^2_Z)]$ 
multiplying the Born amplitude, in agreement with the
general rule obtained in Ref.\cite{Melles,DenPo}
and that the correction to the
$|\Delta\lambda|=0$ amplitude vanishes.\\

A similar set of corrections 
would be provided by the U.V. photon sector (cutted at scale $M_Z$),
just replacing the internal $Z$ by an internal $\gamma$ in all the
diagrams of Fig.1a. The result is given in eq.(\ref{gZm}) and
(\ref{gZp}) (terms with $Q^2_f$ instead 
of ${[g^Z_{Vf}-g^Z_{Af}(2\tau)]^2
/ 4s^2_Wc^2_W}$). The properties of this "$\gamma$ sector"
are exactly similar to those of the $Z$ one. In the following
numerical discussions we shall omit it, taking the stand point that
all photonic corrections (the U.V. ones, 
the I.R. ones, including soft photon emission)
should be put altogether inside 
"QED-type" of corrections which depend on the characteristics
of the detectors and are generally treated separately by 
specific programs. This is obviously a matter of choice, which can
easily be modified.\\

\underline{$W$ sector}\\

The corresponding diagrams are listed in Fig.1b. In addition to those
which are obtained just by replacing the $Z$ by the $W$, 
there now appears new
triangle and box diagrams involving the three-boson $\gamma WW$
coupling.
The resulting amplitudes are given in eqs.(\ref{Wm}),(\ref{Wp}).
One sees that the leading terms eqs.(\ref{Wmlt}),(\ref{Wplt})
are enriched by angular dependent and angular independent contributions
arising from the $\gamma WW$ coupling, which appear in addition to the 
$[\ln^2(s/ M^2_Z)-3\ln (s/ M^2_Z)]$ correction of the 
$|\Delta\lambda|=2$ amplitude.\\

\underline{Higgs sector}\\

In SM the Higgs sector consists in the set of diagrams of Fig.1c
involving charged and neutral $\Phi^{\pm,0}$ Goldstone bosons
as well as the physical $H$ Higgs boson, coupled to fermions
through Yukawa terms proportional to $m_f/M_W$.
This set of diagrams is relevant only for top and
bottom quark production. The resulting amplitudes are given in 
eqs.(\ref{Hm}),(\ref{Hp}) and their leading parts in
eqs.(\ref{Hmlt}),(\ref{Hplt}). As expected from the general
properties established in  \cite{Melles,DenPo},
these leading corrections coming from field renormalization constants
(that one can directly obtain by solely considering external
self-energy contributions) are angular independent, linearly logarithmic
and only affect the
$|\Delta\lambda|=2$ (Born) amplitude.\\

{\bf c)  SUSY additional contributions}\\

In the case of the MSSM, one should add to the previous SM terms
the following additional SUSY corrections. We have separated
them in two parts; first, a "non massive part"
arising from the diagrams of
Fig.2a, in which only the mass-independent parts of the chargino
and neutralino couplings are considered (corresponding to the charged
or neutral "gaugino" components); secondly, a "massive part"
due to the
mass-dependent terms of the chargino and neutralino couplings 
(corresponding to the charged or neutral "higgsino" components)
and also to the diagrams involving SUSY Higgs bosons (to this
last contributions we have subtracted the contribution
of the standard
$H_{SM}$ diagrams in order to not make double counting of the physical
Higgs sector). A general remark, which was already made in 
the case of $e^+e^-$
collisions, is that,
in the asymptotic regime $s>>M^2$, the only dependence in
the MSSM parameters which remains is the dependence in $\tan\beta$; 
all other
parameters (except the global SUSY scale $M$ appearing in the
logarithmic terms) have disappeared because of the unitarity
properties of the mixing matrices appearing in the SUSY couplings,
see the fourth paper of Ref.\cite{alllog}.\\

\underline{Non massive terms}\\

The amplitudes resulting from the mass-independent part of the
diagrams of Fig.2a are given in eqs.(\ref{Sm}),(\ref{Sp}),
and their leading terms in eqs.(\ref{Smlt}),(\ref{Splt}).
For the same reason as in the case of the Higgs sector,
the correction to the
$|\Delta\lambda|=2$ amplitude is only linearly logarithmic, angular
independent (they can also be obtained from the external self-energy
contributions to the field renormalization constants),
and the correction to the
$|\Delta\lambda|=0$ amplitude vanishes asymptotically.\\

\underline{Massive terms}\\

The amplitudes resulting from the mass-dependent part of the
diagrams of Fig.2a and of Fig.2b are given in
eqs.(\ref{SYm}),(\ref{SYp}),
and their leading terms in eqs.(\ref{SYmlt}),(\ref{SYplt}).
They behave asymptotically
in a way similar to the SM Yukawa terms, 
the correction to the
$|\Delta\lambda|=2$ amplitude being also only linearly logarithmic and
angular independent,
and the correction to the
$|\Delta\lambda|=0$ amplitude vanishing. However, an important
fact is the appearance of a $\cot^2\beta$ dependence in the
term proportional to $m^2_t/M^2_W$, and a
$\tan^2\beta$ dependence in the
term proportional to $m^2_b/M^2_W$ (which can be very
important for large  $\tan\beta$ values).\\
We also note that, in the MSSM, summing the SM and the additional SUSY
contributions, the leading 
asymptotic massive terms combine in order
to reproduce the massive SM contributions
in which the $m^2_t/ M^2_W$ terms have been multiplied by
$2(1+\cot^2\beta)$ and the $m^2_b/ M^2_W$ terms by
$2(1+\tan^2\beta)$. This rule had already been obtained for the
process $e^+e^-\to f\bar f$ in the fifth paper of Ref.\cite{alllog}. \\

Let us finish this section 
by making a comparison with
the asymptotic properties observed in the case of
$e^+e^-\to f\bar f$. In the 't Hooft $\xi=1$ gauge, the contributions
of the triangle and box contributions behave sometimes
differently in the $e^+e^-$ and in the $\gamma\gamma$ cases.
The single $Z$ and $W$ triangles get only linear logarithms in
the $\gamma\gamma$ case, whereas they get linear and quadratic
logarithms in $e^+e^-$; on the opposite the $WW$ triangle gets only
a quadratic logarithm in $\gamma\gamma$ instead of the
linear logarithm in $e^+e^-$. These differences are complemented by
those of the box diagrams. In both $Z$ and $W$ sectors, the
boxes produce linear and quadratic logarithms in $\gamma\gamma$,
whereas in the $e^+e^-$ case the $ZZ$ box give only linear
logarithms and the $WW$ box has both linear and quadratic logarithms.
The Higgs and the SUSY sectors are very similar in the
$\gamma\gamma$ and in the $e^+e^-$ cases. They only give linear 
logarithms, arising only from the triangle diagrams (and also from the
internal fermion
self-energy in the $\gamma\gamma$ case). The Higgs and SUSY
box diagrams give no leading logarithms at all, in both
$\gamma\gamma$ and $e^+e^-$ cases.\\

\section{Effects on the $\gamma\gamma\to 
\lowercase{f\bar f}$ observables}

Having obtained
the explicit expressions of the helicity amplitudes, it is easy
to compute the various elements of the polarized $\gamma\gamma$
cross section. The general expression is given in Appendix C.
Due to Bose statistics, CP-invariance and real (asymptotic) amplitudes,
the expression of the cross section including terms
up to order $\alpha^3$ simplifies to:

\bqa
{d\sigma\over d\tau d\cos\theta}&=&{d \bar L_{\gamma\gamma}\over
d\tau} \Bigg \{[1-\langle \xi_2 \xi^\prime_2
\rangle]
{d\bar{\sigma}_0\over d\cos\theta}
+[\langle \xi_2 \rangle-\langle \xi^\prime _2
\rangle]
{d\bar{\sigma}_{2}\over d\cos\theta}
\nonumber\\
&&
+[\langle\xi_3\rangle \cos2\phi +
\langle\xi_3^ \prime\rangle\cos2\phi^\prime]
{d\bar{\sigma}_{3}\over d\cos\theta}
\nonumber\\
&&+[\langle\xi_3 \xi_3^\prime\rangle\cos2(\phi+\phi^\prime)]
{d\bar{\sigma}_{33}\over d\cos\theta}
\nonumber\\
&&+ [\langle\xi_2 \xi_3^\prime\rangle\cos2 \phi^\prime
-\langle\xi_3 \xi^\prime_2\rangle\cos2\phi]
{d\bar{\sigma}_{23}\over d\cos\theta}
 \Bigg \} \ \ ,
\label{sigpols}
\eqa
\noindent
in which $ d \bar L_{\gamma\gamma}/d\tau$ describes the photon-photon
luminosity per unit $e^-e^+$ flux obtained by the laser backscattering
method \cite{laser}; $\tau=s /s_{ee}$ where
$s\equiv s_{\gamma \gamma}$. The Stokes parameters
$(\xi_2, \xi^\prime_2)$, $(\xi_3,~ \xi^\prime_3)$
and $(\phi,~ \phi^\prime)$ describe respectively
 the average helicities, transverse
polarizations and azimuthal angles of the two
backscattered photons, see ref.\cite{ggsig},

The Born amplitudes only feed the (Parity conserving)
${d\bar{\sigma}_0/ d\cos\theta}$
and ${d\bar{\sigma}_{33}/ d\cos\theta}$ terms.
The one-loop effects feed all the above terms. Note the specific
photon polarization dependences which can be used to test
the structure of the one-loop electroweak corrections and the absence
of unexpected effects.
Taking into account the
fact that

$${d\bar \sigma_0\over d\cos\theta},~~~
{d\bar \sigma_3\over d\cos\theta},~~~
{d\bar \sigma_{33}\over d\cos\theta} 
$$
are \underline{$\cos\theta$-symmetric} and
$${d\bar \sigma_2\over d\cos\theta},~~~
{d\bar \sigma_{23}\over d\cos\theta}$$
\underline{$\cos\theta$-antisymmetric},
we construct the \underline{five} ratios;\\

\bq
R_0=\int d\cos\theta~ [{d\bar \sigma_0\over d\cos\theta}-
{d\bar \sigma^{Born}_0\over d\cos\theta}]~/\int d\cos\theta~
{d\bar \sigma^{Born}_0\over d\cos\theta}
\label{R0s}\eq

\bq
R_{33}=\int d\cos\theta~ [{d\bar \sigma_{33}\over d\cos\theta}-
{d\bar \sigma^{Born}_{33}\over d\cos\theta}]~/\int d\cos\theta~
{d\bar \sigma^{Born}_0\over d\cos\theta}
\label{R33s}\eq

\bq
R_2=\int_{F-B} d\cos\theta~ [{d\bar \sigma_2\over d\cos\theta}]
~/\int d\cos\theta~
{d\bar \sigma^{Born}_0\over d\cos\theta}
\label{R2s}\eq

\bq
R_3=\int d\cos\theta~ [{d\bar \sigma_3\over d\cos\theta}]
~/\int d\cos\theta~
{d\bar \sigma^{Born}_0\over d\cos\theta}
\label{R3s}\eq

\bq
R_{23}=\int_{F-B} d\cos\theta~ [{d\bar \sigma_{23}\over d\cos\theta}]
~/\int d\cos\theta~
{d\bar \sigma^{Born}_0\over d\cos\theta}
\label{R23s}\eq
\noindent
on which the electroweak effects are now illustrated and
discussed.\\

One should first note, using the definitions of the various "cross
sections" given in Appendix C, that the last two ratios $R_3$ and
$R_{23}$ only involve products of $|\Delta\lambda=0|$ with
$|\Delta\lambda=2|$ amplitudes. As we have seen that, in the
asymptotic regime (see for example the leading expressions written in
Appendix A), the one-loop contributions to
$|\Delta\lambda=0|$ amplitudes are much weaker than the one to
$|\Delta\lambda=2|$ amplitudes, one expects that these two ratios
are much weaker than the other three ones.\\

{\bf Angular distributions}\\

 The angular distribution of the unpolarized Born cross section
${d\sigma^{Born}_0/ d\cos\theta}$ 
is (symmetrically) strongly peaked in the forward
and backward directions, see Fig.3a-c at $3~TeV$.
The electroweak corrections 
modify somewhat this distribution because 
their effect is larger in the central
region, as shown in Fig.4a-c where we plot 
the angular dependence of the
relative effect of the electroweak corrections, defined as

\bq
\Delta({d\sigma_0\over d\cos\theta})\equiv
[{d\sigma_0\over d\cos\theta}-{d\sigma^{Born}_0\over d\cos\theta}]
/{d\sigma^{Born}_0\over d\cos\theta}\ .
\eq
\noindent
It will therefore be interesting
to have the largest possible angular acceptance allowed by
experimental detection and to cut the angular distribution
into several bins. One could then check the relative 
increase of the weak
corrections in the central region.
\par
Note that the radiative correction effect is always negative,
that the supersymmetric corrections always increase the magnitude
of the effect,
and in the case of $t\bar t$, $b\bar b$ that this effect 
strongly depends on $\tan\beta$.\\

We now study in more details the behaviour of these effects
versus the energy, by considering the integrated
cross sections. In the following illustrations
we choose to integrate the angular
distributions in the domain $30^0<\theta<150^0$.\\

{\bf Leading versus non leading terms}\\

It is interesting to compare, as a function of the energy,
the relative importance of the various logarithmic terms
which have been presented in the previous Section II.
We will do that by considering the ratio $R_0$ giving
the relative electroweak effects on the unpolarized
cross section, defined in eq.(\ref{R0s}).\par 
In Fig.5a,b for $l^+l^-$, Fig.6a,b,c for $t\bar t$,
Fig.7a,b,c for $b\bar b$ we show, separately for the SM and the MSSM
cases, the contribution
of the sum of all logarithmic terms
(collected in Appendix A and B), compared to the results obtained when
dropping the non leading logarithmic terms (i.e. terms of the type
$\ln^2(t/s)$,... etc) and also to the results
obtained when dropping, in addition, 
the leading angular dependent terms (terms $ln(s/M^2)$ multiplied
by angular dependent logarithms).\par
One sees that the non leading logarithmic terms 
(which appear in the expressions of the box
contributions given in Appendix B) behave roughly like an additional
small constant contribution (of the order of one percent)
whose relative importance as compared to the full
electroweak correction, decreases with the energy; this is true for
both the SM and the MSSM cases.\par
On the contrary, the leading angular dependent terms (which appear
in the triangle diagrams involving the $\gamma WW$ three boson
coupling) are more important (similar effects has been noticed in
Ref.\cite{DenPo}, in the case of the crossed channel
$e^+e^-\to\gamma\gamma$) and increase with the energy. They cannot
be omitted at all, and we will come back to their role in the final
discussion. This comment applies to both
the SM and the MSSM cases, as the SUSY additional
contributions only consist 
in angular independent contributions.\\

We have checked that, around $1$ TeV, our asymptotic results agree with
those obtained in Ref.\cite{dendit} for the purely weak part of the
SM corrections to light fermion pair production. In the $t\bar t$ case,
the agreement at $1$ TeV is only qualitative, for both the SM case
\cite{dendit} and the MSSM case \cite{Zhou}, as this energy is just
marginally "asymptotic" for top quarks and for supersymmetric
contributions. Nevertheless the cancellation of the various MSSM
parameters, except for the large $\tan\beta$ dependence that we
emphasized, can already be seen at this energy in \cite{Zhou}.\\

{\bf Importance of Yukawa terms}\\

In Fig.6a,b,c for $t\bar t$,
Fig.7a,b,c for $b\bar b$, we have also shown the effect of
dropping the Yukawa terms (coming from the Higgs and the Higgsino
sectors) proportional to $m^2_t/M^2_W$ and $m^2_b/M^2_W$.
Comparing the curves for the SM case and the curves for the case with
no Yukawa terms in Fig.6a for 
$t\bar t$, and Fig.7a for $b\bar b$, one sees that these terms are very
important, especially in the $t\bar t$ case, where they contribute
easily for half of the effect at CLIC energies.
In the MSSM case, the comparison is made in Fig.6b and 7b for
$\tan\beta=4$, and in Fig.6c and 7c for
$\tan\beta=40$. The $\tan\beta$ dependence can be understood by
looking at eq.(\ref{SYmlt}), in which one sees
a $\cot^2\beta$ dependence associated to the term $(m^2_t/M^2_W)$
and a $\tan^2\beta$ dependence associated to the $(m^2_b/M^2_W)$,
which becomes dominant at very large $\tan\beta$ values.
These properties are rather similar to those observed
in $\gamma\gamma\to f \bar f$ \cite{tgbeta}.\\

{\bf Polarized and unpolarized cross sections versus the energy}\\

We finally illustrate the behaviour of the various terms of the
polarized cross section, eq.(\ref{sigpols}), versus the energy, in
the $l^+ l^-$, $t\bar t$ and $b\bar b$ cases.\par
In Fig.8a,b,c and 9a,b,c we present the ratios $R_0$ and $R_{33}$
which show the relative departures from the Born prediction,
see eq.(\ref{R0s}),(\ref{R33s}).
The effects are in all cases of the order of
several percents at LC energies and of the order of 10-20
percents at CLIC energies. In the MSSM case they are larger than in the
SM case, especially for large $\tan\beta$ values.\par
In Fig.10a,b,c , 11a,b,c and 12a,b,c we present the ratios 
$R_2$, $R_3$ and $R_{23}$ defined
in eqs.(\ref{R2s}),(\ref{R3s}),(\ref{R23s}). There is no 
Born contribution to these terms.
The effects in $R_2$ (circular photon polarization dependence)
are comparable to those previously seen in
$R_0$. This is because $R_2$ measures the Parity violating effects
which are maximal in $W$ couplings. On the contrary, the effects
are very small in $R_3$ (one photon transversally polarized)
and $R_{23}$ (one photon transversally polarized, the other one being
circularly polarized)) because these terms, as we have already
mentioned after their definitions, are proportional to the
interference of small $\Delta\lambda=0$ amplitudes (which have no
leading $ln(s/M^2)$ or $ln^2(s/M^2)$ terms) with
$\Delta\lambda=2$ ones. Very high energies are required in order for
these observables to reach
the observable percent level.\\

We can add a final remark concerning the cross section for
$\gamma\gamma$ to hadrons, the analogue 
of the cross section for hadron production in $e^+e^-$ collisions
$\sigma_5\equiv\sigma(e^+e^-\to u\bar u+c\bar c + d\bar d +
s\bar s + b\bar b)$. 
In $\gamma\gamma$ collisions, as we can see
from Fig.3b,c , because of the factor $Q_f^4$ in the Born cross
section, the rate is largely dominated the contribution of
up-quarks ($u,c$), and the Yukawa contribution, appearing solely
in the $b$ case, can be completely neglected. So the properties
of the electroweak radiative corrections to
$\sigma(\gamma\gamma \to hadrons)$ can be totally inferred
from those of $\sigma(\gamma\gamma \to t\bar t)$, ignoring the
Yukawa contributions; see for example the curves corresponding to the
case with no Yukawa terms in Fig.6ab.\\

\section{Conclusions}

We have studied the high energy behaviour of the one-loop weak
corrections to the processes $\gamma\gamma\to f \bar f$, 
in SM and in MSSM.\\

In the asymptotic energy regime, we have classified 
and computed all correction terms coming in the 't Hooft $\xi=1$
gauge, from fermion self-energies, triangle and box diagrams.
We have checked that, in each weak sector, the set of diagrams
contributes in a
gauge-independent way to the linear and quadratic logarithmic
contributions to the $\gamma\gamma\to f \bar f$ amplitudes.
Explicit analytic expressions are given in
Appendix A and B, and turn out to be rather simple, 
and reflecting in a remarkable way the
theoretical properties of the SM charged gauge,
neutral gauge and
Higgs sectors and of the MSSM gaugino and Higgsino sectors.
These results satisfy the known general properties 
of leading electroweak logarithms at one loop \cite{KPS,Melles,DenPo}.
They also match with the complete one loop computations performed 
around 1 TeV in\cite{dendit,Zhou}.\\

We have shown that these effects
should be well visible in $\gamma\gamma$
collisions at LC and CLIC, the large luminosities expected
at these machines allowing to reach an accuracy better 
than the percent level.
We have given the results for
five observables defined in the case of polarized photon beams. 
Clearly, the behaviour of each observable should provide 
clean tests of the SM or the 
MSSM and allow to check the absence of unexpected new physics effect.\\

An important fact is the strong rise of the effect on the cross section,
partly due to the angular independent factor ${\alpha/4\pi}
[\ln^2(s/ M^2_W)-3\ln(s/ M^2_W)]$, but we have shown that
there are also important angular dependent contributions.
A clear difference also appear, in each $f=l,~t,~b$ case, 
between the SM and the MSSM corrections. 
The SUSY additional terms increase the magnitude of
the weak corrections. For example at $3$ TeV, in $l^+l^-$ production,
the correction is -12.7\% in SM and -13.6\% in MSSM.
In the $t\bar t$ and $b\bar b$ cases,
the Yukawa terms contribute
for a large part of the effects, both in SM and in MSSM; in this last
case an observable $\tan\beta$ dependence appear. At $3$ TeV,
the weak effects to $t\bar t$ production are -23.1\% in SM,
-27.2\% in MSSM($\tan\beta=4$), -28.6\% in MSSM($\tan\beta=40$);
and for $b\bar b$ production, they are -32.3\% in SM,
-34.8\% in MSSM($\tan\beta=4$), -41.6\% in MSSM($\tan\beta=40$).
This $\tan\beta$ dependence could 
be used for a $\tan\beta$ measurement (see the corresponding
discussion in $e^+e^-$ collisions in Ref.\cite{tgbeta}).\\

These results are complementary to those observed in the process
$e^+e^-\to f \bar f$. We have shown that the role of the
self-energy, triangle and box diagrams are different in the two
processes, but the qualitative aspect of the information
that can be reached about the features of the electroweak corrections
is rather similar. There are however quantitative differences
when comparing the effects in $l^+l^-$, $b\bar b$ and $t\bar t$
production. This is essentially due to the fact that in
$\gamma\gamma$ collisions the Born term,
proportional to $Q^4_f$, is especially small in the $b\bar b$ case, so
that the electroweak corrections are relatively larger.
Also the effects
of gauge, Yukawa, and SUSY contributions cumulate so that
the corrections are larger 
than in the $e^+e^-\to f \bar f$ processes
at the same energy.\\

As these first order effects already reach the 10 percent 
level around 1 TeV,
and 30 percent around 3 TeV, one may naively expect that 
higher order terms easily reach the few percent level,
observable at CLIC, 
raising the question of a possible two-loop computation.
For the angular independent terms, general resummation techniques
has been proposed \cite{resum}, which would partly solve the
problem. However we have shown that there are important angular
dependent terms for which no prescription has yet been obtained
and may require an explicit two-loop computation.\\

At lower energies (the 0.5 to 1 TeV domain of LC), there is
apparently no such problem. Although the effect in
$\gamma\gamma \to b\bar b$ can reach 15 percent at 1 TeV,
the weaker experimental accuracy in this channel, may still
allow to stay at the one-loop level. However, as we have shown
by comparing leading and non leading logarithmic terms, in this
energy range, the logarithmic approximation is probably not
sufficient. Constant terms (and possibly terms of order $M^2/s$)
may not be negligible, especially if the SUSY scale is rather
high and one may not be allowed to neglect 
the mass of the SUSY particles
running inside the loops. This approximation also fails
to reproduce the "resonance" effects which appear
around the thresholds for (sfermion or chargino) pair
production\cite{Zhou}. In this "low energy" regime, the full set of
MSSM parameters enter the game (and not only $\tan\beta$
as in the asymptotic regime). We intend to perform
a detailed comparison of the
logarithmic approximation with the exact computation of the
full one-loop contributions.
It should allow to understand the role and to discuss the
measurability, in the
LC regime, of each of the various MSSM parameters.

\newpage

\appendix

\section{Asymptotic expressions of the helicity amplitudes at one-loop}

We denote by $F_{\lambda,\lambda',\tau,\tau'}$ the helicity
amplitudes of the process $\gamma\gamma \to f \bar f$, 
$\lambda,\lambda',\tau,\tau'$ being the helicities of the
photons ($\pm1$), and of the fermion and antifermion ($\pm{1/2}$)
in the $\gamma\gamma$ center of mass. We denote by
$e,l$, $e',l'$ the photon polarization
vectors and 4-momenta and $p,p'$ the fermion, antifermion 
4-momenta; $q=p-l=l'-p'$, $q'=p-l'=l-p'$;
 $\sqrt{s}$, $\theta$ are the 
energy and the scattering angle.\par
We work in the high energy limit 
$s=(l+l')^2=(p+p')^2,~t=q^2=-~{s/2}(1-\cos\theta),~u=q'^2
=-~{s/2}(1+\cos\theta)>> M^2$ 
(avoiding the forward and backward domains),
keeping only logarithmic terms involving $s$, $t$ or $u$.
A general consequence of the high energy limit is the dominance
of chirality conserving terms with $\tau'=-\tau$ only.\\

{\bf a) Born term}\\

At high energy, the invariant amplitude
corresponding to the diagrams of Fig.1 is:

\bq
{\cal R}^{Born} = -~e^2 Q^2_f \bar u_f(p)[{ e\!\!\! / \  
q \!\!\! / \  e \!\!\! / \ '
\over t}+{e\!\!\! / \ '  q\!\!\! / \ '  e\!\!\! / \ \over u}]
v_{\bar f}(p')
\eq 
\noindent
$Q_f$ is the fermion charge in unit of $|e|$.

It
leads to the helicity amplitudes

\bq F^{Born}_{\lambda,-\lambda,\tau,-\tau}=-~8\pi\alpha
Q^2_f[{\lambda+2\tau \cos\theta\over \sin\theta}]
\eq

Note that, at high energy, due to Bose symmetry, the Born term
only involves $\lambda'=-\lambda$ (i.e.
$|\Delta\lambda|=2$) amplitudes.\\

{\bf b) SM  electroweak corrections}\\

{\bf $\gamma$ and $Z$ sector}\\

The sum of self-energy, triangle and Box diagrams of Fig.1a (to which
external fermion self-energy diagrams are added)
is convergent and gives the asymptotic contributions:

\bqa F_{\lambda,-\lambda,\tau,-\tau}&=&
\alpha^2Q^2_f\{Q^2_f+{[g^Z_{Vf}-g^Z_{Af}(2\tau)]^2
\over 4s^2_Wc^2_W}\}\{~2~[{\lambda+(2\tau) \cos\theta
\over \sin\theta}]~\ln {s\over M^2_Z} + B^1_{\lambda,-\lambda}(M^2_Z)\}
\label{gZm}\eqa
\bqa F_{\lambda,\lambda,\tau,-\tau}&=&
\alpha^2Q^2_f\{Q^2_f+{[g^Z_{Vf}-g^Z_{Af}(2\tau)]^2
\over 4s^2_Wc^2_W}\}\{-8~[{(2\tau) \cos\theta
\over \sin\theta}]~\ln {s\over M^2_Z} + B^1_{\lambda,\lambda}(M^2_Z)\}
\label{gZp}\eqa
\noindent
The box quantities
$B^i$ are defined in Appendix B, and 
$g^Z_{Vf}=I_{3f}(1-4s^2_W|Q_f|)$, $g^Z_{Af}=I_{3f}$.\\

\vspace{4cm}

\underline{leading terms}
\bqa
 F^{l.t.}_{\lambda,-\lambda,\tau,-\tau}&&
\to -
F^{Born}_{\lambda,-\lambda,\tau,-\tau}
({\alpha\over4\pi})\{Q^2_f+{[g^Z_{Vf}-g^Z_{Af}(2\tau)]^2
\over 4s^2_Wc^2_W}\}(\ln^2{s\over M^2_Z}-3\ln {s\over M^2_Z})
\label{gZmlt}\eqa

\bqa F^{l.t.}_{\lambda,\lambda,\tau,-\tau}&&\to 0
\label{gZplt}\eqa

{\bf $W$ sector}\\

We now sum the contributions of the charged gauge sector, with
the self-energy, triangle and box diagrams of Fig.1b. Note that
in order to get a convergent result, one has to add the photon
self-energy contribution; it cancels
the divergent contribution which appears in
the axial term of 
the corrected $\gamma ff$ vertex; whereas 
a remaining divergence in the vector term is absorbed by 
the charge renormalization.

\bqa F_{\lambda,-\lambda,\tau,-\tau}&=& 
{\alpha^2\over4s^2_W}[1-(2\tau)]\{
[{\lambda+(2\tau) \cos\theta\over \sin\theta}]
[2Q_f(Q_f-2(2I_{3f}))\ln {s\over M^2_W}\nonumber\\
&&
+Q_f(2I_{3f})[(1+\cos\theta)\ln^2{t\over m^2_W}+
(1-\cos\theta)\ln^2{u\over m^2_W}]
+2\sin^2\theta~\ln {s\over M^2_W}]\nonumber\\
&&+[Q_{f}-(2I_{3f})]^2B^1_{\lambda,-\lambda}(M^2_W)
+B^2_{\lambda,-\lambda}(M^2_W)
-[Q_{f}-(2I_{3f})](2I_{3f})B^5_{\lambda,-\lambda}(M^2_W)\}
\label{Wm}\eqa

\bqa F_{\lambda,\lambda,\tau,-\tau}&=& 
{\alpha^2\over4s^2_W}[1-(2\tau)]\{(2\tau)[
-8Q_f(Q_{f}-(2I_{3f})){\cos\theta\over \sin\theta}
\ln {s\over M^2_W}\nonumber\\
&&+Q_f(2I_{3f})\sin\theta[
{(2-\cos\theta)\over1-\cos\theta}\ln^2{t\over m^2_W}-
{(2+\cos\theta)\over1+\cos\theta}\ln^2{u\over m^2_W}]\nonumber\\
&&-2\sin\theta \cos\theta)\ln {s\over M^2_W}]\nonumber\\
&&+[Q_{f}-(2I_{3f})]^2B^1_{\lambda,\lambda}(M^2_W)
+B^2_{\lambda,\lambda}(M^2_W)
-[Q_{f}-(2I_{3f})](2I_{3f})B^5_{\lambda,\lambda}(M^2_W)\}
\label{Wp}\eqa

\underline{leading terms}

\bqa
 F^{l.t.}_{\lambda,-\lambda,\tau,-\tau}&&
\to
-F^{Born}_{\lambda,-\lambda,\tau,-\tau}({\alpha\over32\pi
s^2_W Q^2_f})[1-(2\tau)]
\{2Q^2_f(\ln^2{s\over M^2_W}-3\ln {s\over M^2_W})+4Q_f(2I_{3f})
\ln^2{s\over M^2_W}\nonumber\\
&&+4[\cos\theta \ln{1-\cos\theta\over1+\cos\theta}
+(2Q_f(2I_{3f})-1)\ln{\sin^2\theta\over4}]\ln {s\over M^2_W}\}
\label{Wmlt}\eqa

\bqa F^{l.t.}_{\lambda,\lambda,\tau,-\tau}&&
\to 0
\label{Wplt}\eqa

Note the appearance of angular dependent leading terms. This is the
only sector where it happens (such terms were also found in
Ref.\cite{DenPo} in the crossed channel $e^+e^-\to \gamma\gamma$ for
left-handed electrons). See the discussion in Sections II and
III.\\

\vspace{4cm}

{\bf Higgs sector}

We now add the contributions of the diagrams of Fig.1c involving the
Goldstone $\Phi$ and the physical Higgs $H_{SM}$.
This concerns only the production of massive quarks $f=t,b$, 
as these contributions, arising from the Yukawa couplings, are
proportional to $m^2_f/M^2_W$.

\bqa F_{\lambda,-\lambda,\tau,-\tau}&=& {\alpha^2\over4s^2_W}\{
Q^2_f
[{\lambda+(2\tau) \cos\theta\over \sin\theta}].\nonumber\\
&&
.\{[{m^2_t\over M^2_W}(1+(2\tau)(2I_{3f}))
+{m^2_b\over M^2_W}(1-(2\tau)(2I_{3f}))]+2({m^2_f\over M^2_W})\}
\ln{s\over M^2_W}\nonumber\\
&&+[B^4_{\lambda,-\lambda}+(Q_{f}-(2I_{3f}))^2B^3_{\lambda,-\lambda}
-(Q_{f}-(2I_{3f}))(2I_{3f})B^6_{\lambda,-\lambda}].\nonumber\\
&&.[{m^2_t\over M^2_W}(1+(2\tau)(2I_{3f}))
+{m^2_b\over M^2_W}(1-(2\tau)(2I_{3f}))]
+2Q_f^2({m^2_f\over M^2_W})B^3_{\lambda,-\lambda}\}\
\label{Hm}\eqa

\bqa F_{\lambda,\lambda,\tau,-\tau}&=& {\alpha^2\over4s^2_W}\{
[B^4_{\lambda,\lambda}+(Q_{f}-(2I_{3f}))^2B^3_{\lambda,\lambda}
-(Q_{f}-(2I_{3f}))(2I_{3f})B^6_{\lambda,\lambda}].\nonumber\\
&&.[{m^2_t\over M^2_W}(1+(2\tau)(2I_{3f}))
+{m^2_b\over M^2_W}(1-(2\tau)(2I_{3f}))]
+2Q_f^2({m^2_f\over M^2_W})B^3_{\lambda,\lambda}\}
\label{Hp}\eqa

\underline{leading terms}

\bqa F^{l.t.}_{\lambda,-\lambda,\tau,-\tau}&&
\to -F^{Born}_{\lambda,-\lambda,\tau,-\tau}({\alpha
\over32\pi s^2_W})\{
[{m^2_t\over M^2_W}(3+(2\tau))
+{m^2_b\over M^2_W}(1-(2\tau))]\delta_{tf}\nonumber\\
&&+ [{m^2_b\over M^2_W}(3+(2\tau))
+{m^2_t\over M^2_W}(1-(2\tau))]\delta_{bf}\}\ln{s\over M^2_W}
\label{Hmlt}\eqa

\bqa F^{l.t.}_{\lambda,\lambda,\tau,-\tau}&
\to 0
\label{Hplt}\eqa

Note that the box functions $B^{3,4,6}$ (and consequently, the full
Higgs contribution to $F_{\lambda,\lambda,\tau,-\tau})$
do not contribute to the leading 
$\ln s$ or $\ln^2s$ terms; so no scale is mentioned in their notation
(see Appendix B); the same property holds in the following 
supersymmetric contributions.\\

{\bf c)  SUSY additional contributions}\\

{\bf Non massive terms}\\

By non massive terms we mean the contributions
due to the diagrams involving gauge couplings of
sfermions, charginos and neutralinos.
They come from self-energy, triangle and Box diagrams in Fig.2a (and
external fermion self-energy terms).

\bqa F_{\lambda,-\lambda,\tau,-\tau}&=& {\alpha^2\over4s^2_W}\{
[{\lambda+(2\tau) \cos\theta\over \sin\theta}][
2Q^2_f({2 C_fQ^2_f\over c^2_W}+1-(2\tau))]\ln {s\over M^2}\nonumber\\
&&+[1-(2\tau)][2B^3_{\lambda,-\lambda}-2(Q_{f}-(2I_{3f}))(2I_{3f})
B^6_{\lambda,-\lambda}
+D_fB^4_{\lambda,-\lambda}]\nonumber\\
&&+[1+(2\tau)]E_fB^4_{\lambda,-\lambda}\}
\label{Sm}\eqa

\bqa F_{\lambda,\lambda,\tau,-\tau}&=& {\alpha^2\over4s^2_W}\{
[1-(2\tau)][2B^3_{\lambda,\lambda}-2(Q_{f}-(2I_{3f}))(2I_{3f})
B^6_{\lambda,\lambda}
+D_fB^4_{\lambda,\lambda}]\nonumber\\
&&+[1+(2\tau)]E_fB^4_{\lambda,\lambda}\}
\label{Sp}\eqa
\noindent
where\\
$$C_l={1\over4}(1-(2\tau))+s^2_W(1+(2\tau)),~~~
D_l={1\over c^2_W},~~~E_l={4s^2_W\over c^2_W}\ ,$$

$$C_t={9-8s^2_W\over36}(1-(2\tau))+{4s^2_W\over9}(1+(2\tau)),~~~
D_t={2\over9}+{4(9-8s^2_W)\over81c^2_W},~~~
E_t={64s^2_W\over81c^2_W}\ ,$$

$$C_b={9-8s^2_W\over36}(1-(2\tau))+{s^2_W\over9}(1+(2\tau)),~~~
D_b={8\over9}+{(9-8s^2_W)\over81c^2_W},~~~
E_b={4s^2_W\over81c^2_W}\ .$$

\underline{leading terms}

\bqa F^{l.t.}_{\lambda,-\lambda,\tau,-\tau}&&
\to -F^{Born}_{\lambda,-\lambda,\tau,-\tau}({\alpha
\over16\pi s^2_W})[({2 C_fQ^2_f\over c^2_W}+1-(2\tau))]\ln {s\over M^2}
\label{Smlt}\eqa

\bqa F_{\lambda,\lambda,\tau,-\tau}&&\to 0
\label{Splt}\eqa

$M$ is a common SUSY scale introduced for convenience
(that will be fixed to $0.5~TeV$ in the
illustrations). Note that a change of value of $M$ amounts to
the introduction of additional (neglected) constant terms, as the
SUSY contributions only appear with $\ln(s/ M^2)$ and never
with quadratic logarithmic terms.\\
Note in addition that the SUSY contribution 
to $F_{\lambda,\lambda,\tau,-\tau}$
has also no leading $\ln s$ or $\ln^2s$ term.

{\bf Massive terms}

These terms arise from the Yukawa couplings of the Higgsino
component of the charginos and neutralinos interacting with sfermions,
as well as from the physical SUSY Higgs contributions (from which
we subtract the SM Higgs contribution in order to not make
double counting of the Higgs sector contribution). From self-energy, 
triangle and box diagrams of Fig.2a,b (and external
fermion self-energy terms) one
gets:

\bqa F_{\lambda,-\lambda,\tau,-\tau}&=& {\alpha^2\over4s^2_W}\{
Q^2_f
[{\lambda+(2\tau) \cos\theta\over \sin\theta}].\nonumber\\
&&\{
.[{m^2_t\over M^2_W}(1+(2\tau)(2I_{3f}))(1+2\cot^2\beta)
+{m^2_b\over M^2_W}(1-(2\tau)(2I_{3f}))(1+2\tan^2\beta)]\nonumber\\
&&
+2({m^2_f\over M^2_W})[(1+2\cot^2\beta)\delta_{tf}+
(1+2\tan^2\beta)\delta_{bf}]\}
\ln{s\over M^2}\nonumber\\
&&+(B^4_{\lambda,-\lambda}+(Q_{f}-(2I_{3f}))^2B^3_{\lambda,-\lambda})
[{m^2_t\over M^2_W}(1+(2\tau)(2I_{3f}))\cot^2\beta\nonumber\\
&&
~~~~~~~~~~~~~~~~~~~~~~~~~~~~~~~~~~~~~~~~~~~~~~~
+{m^2_b\over M^2_W}(1-(2\tau)(2I_{3f}))\tan^2\beta]\nonumber\\
&&+(B^3_{\lambda,-\lambda}+(Q_{f}-(2I_{3f}))^2B^4_{\lambda,-\lambda})
[{m^2_t\over M^2_W}(1+(2\tau)(2I_{3f}))(1+\cot^2\beta)\nonumber\\
&&~~~~~~~~~~~~~~~~~~~~~~~~~~~~~~~~~~~~~~~~~~~~~~~
+{m^2_b\over M^2_W}(1-(2\tau)(2I_{3f}))(1+\tan^2\beta)]\nonumber\\
&&-(Q_{f}-(2I_{3f}))(2I_{3f})
B^6_{\lambda,-\lambda}
.[{m^2_t\over M^2_W}(1+(2\tau)(2I_{3f}))(1+2\cot^2\beta)\nonumber\\
&&~~~~~~~~~~~~~~~~~~~~~~~~~~~~~~~~~~~~~~~~~~~~~~~
+{m^2_b\over M^2_W}(1-(2\tau)(2I_{3f}))(1+2\tan^2\beta)]\nonumber\\
&&+2Q^2_f{m^2_f\over M^2_W}
(B^3_{\lambda,-\lambda}[\cot^2\beta\delta_{tf}+
\tan^2\beta\delta_{bf}]+
B^4_{\lambda,-\lambda}[(1+\cot^2\beta)\delta_{tf}+
(1+\tan^2\beta)\delta_{bf}])\}
\label{SYm} 
\eqa

\bqa F_{\lambda,\lambda,\tau,-\tau}&=& {\alpha^2\over4s^2_W}\{
(B^4_{\lambda,\lambda}+(Q_{f}-(2I_{3f}))^2B^3_{\lambda,\lambda})
[{m^2_t\over M^2_W}(1+(2\tau)(2I_{3f}))\cot^2\beta\nonumber\\
&&~~~~~~~~~~~~~~~~~~~~~~~~~~~~~~~~~~~~~~~~~~~~~~~
+{m^2_b\over M^2_W}(1-(2\tau)(2I_{3f}))\tan^2\beta]\nonumber\\
&&+(B^3_{\lambda,\lambda}+(Q_{f}-(2I_{3f}))^2B^4_{\lambda,\lambda})
[{m^2_t\over M^2_W}(1+(2\tau)(2I_{3f}))(1+\cot^2\beta)\nonumber\\
&&~~~~~~~~~~~~~~~~~~~~~~~~~~~~~~~~~~~~~~~~~~~~~~~
+{m^2_b\over M^2_W}(1-(2\tau)(2I_{3f}))(1+\tan^2\beta)]\nonumber\\
&&-(Q_{f}-(2I_{3f}))(2I_{3f})
B^6_{\lambda,\lambda}
.[{m^2_t\over M^2_W}(1+(2\tau)(2I_{3f}))(1+2\cot^2\beta)\nonumber\\
&&~~~~~~~~~~~~~~~~~~~~~~~~~~~~~~~~~~~~~~~~~~~~~~~
+{m^2_b\over M^2_W}(1-(2\tau)(2I_{3f}))(1+2\tan^2\beta)]\nonumber\\
&&+2Q^2_f{m^2_f\over M^2_W}
(B^3_{\lambda,\lambda}[\cot^2\beta\delta_{tf}+
\tan^2\beta\delta_{bf}]+
B^4_{\lambda,\lambda}[(1+\cot^2\beta)\delta_{tf}+
(1+\tan^2\beta)\delta_{bf}])\}
\label{SYp}\eqa

\underline{leading terms}

\bqa F^{l.t.}_{\lambda,-\lambda,\tau,-\tau}&&\to
 -F^{Born}_{\lambda,-\lambda,\tau,-\tau}({\alpha
\over32\pi s^2_W})\{
[{m^2_t\over M^2_W}(3+(2\tau))(1+2\cot^2\beta)
+{m^2_b\over M^2_W}(1-(2\tau))(1+2\tan^2\beta)]\delta_{tf}\nonumber\\
&&+ [{m^2_b\over M^2_W}(3+(2\tau))(1+2\tan^2\beta)
+{m^2_t\over M^2_W}(1-(2\tau))(1+2\cot^2\beta)]\delta_{bf}\}
\ln{s\over M^2}
\label{SYmlt} 
\eqa

\bqa F^{l.t.}_{\lambda,\lambda,\tau,-\tau}&&
\to 0
\label{SYplt}\eqa

\newpage

\section{Asymptotic expressions of the Box diagrams}

The contributions of the Box diagrams of Fig.1,2 to the helicity
amplitudes can be written in the following general form, where
$i=1,...6$ correspond to the $6$ types of Box diagrams.
The following expressions are obtained by retaining only the
logarithmic terms which appear in the complete expressions written in
terms of Passarino-Veltman functions.

\bq
B^i_{\lambda,-\lambda}(M^2)\equiv {s\over2}\sin\theta
[(\lambda+(2\tau) \cos\theta)X^i_1+s~ \sin^2\theta (2\tau) X^i_2]
\eq
\bq
B^i_{\lambda,\lambda}(M^2)\equiv-~{s\over2}\sin\theta(2\tau)
[\cos\theta X^i_1+s~ \sin^2\theta X^i_2+X^i_3]
\eq

\bqa
X^1_1&=&{s^2+tu\over tu^2}\ln^2{t\over s}+{s^2+tu\over ut^2}\ln^2{u\over s}
-{3t+2s\over ut}\ln{t\over s}-{3u+2s\over ut}\ln{u\over s}\nonumber\\
&&
+{t-2s\over ut}\ln{t\over M^2}+{u-2s\over ut}\ln{u\over M^2}
-{4s+5t+5u\over tu}\ln{s\over M^2}+{s\over tu}\ln^2{s\over M^2}
\eqa
\bqa
X^1_2&=&{s-u\over2u^3}\ln^2{t\over s}-{s-t\over2t^3}\ln^2{u\over s}
+{t+3s\over2tu^2}\ln{t\over s}-{u+3s\over2ut^2}\ln{u\over s}
\eqa
\bqa
X^1_3&=&{tu-s^2\over tu^2}\ln^2{t\over s}-{tu-s^2\over ut^2}
\ln^2{u\over s}-{2\over u}\ln{t\over s}+{2\over t}\ln{u\over s}
+{u-t\over tu}\ln^2{s\over M^2}
\eqa

\bqa
X^2_1&=&{s^2+2st\over tu^2}\ln^2{t\over s}+{s^2+2su\over ut^2}
\ln^2{u\over s}
-{3t+4s\over 2ut}\ln{t\over s}-{3u+4s\over 2ut}\ln{u\over s}
-{5\over 2u}\ln{t\over M^2}\nonumber\\
&&
-{5\over 2t}\ln{u\over M^2}
-{13s^2-4t^2-4u^2\over 2stu}\ln{s\over M^2}+{s\over tu}\ln^2{s\over M^2}
-{1\over t}\ln^2{t\over M^2}-{1\over u}\ln^2{u\over M^2}
\eqa
\bqa
X^2_2&=&{t-u\over2u^3}\ln^2{t\over s}-{u-t\over2t^3}\ln^2{u\over s}
+{2t-3u\over2tu^2}\ln{t\over s}-{2u-3t\over2ut^2}\ln{u\over s}
\eqa
\bqa
X^2_3&=&-~{s^3+4t^3+4ts^2+6st^2\over stu^2}\ln^2{t\over s}
+{s^3+4u^3+4us^2+6su^2\over sut^2}\ln^2{u\over s}
+{s+4t\over ut}\ln{t\over s}\nonumber\\
&&
-{s+4u\over ut}\ln{u\over s}+{s\over ut}\ln{t\over u}
+{2t^2+4st-2u^2-4su\over stu}\ln{s\over M^2}\nonumber\\
&&
+({t+2s\over 2st}-{u+2s\over 2su})\ln^2{s\over M^2}+
{4\over s}(\ln^2{t\over M^2}-\ln^2{u\over M^2})
\eqa

\bq
X^3_1={t\over2u^2}\ln^2{t\over s}+{u\over2t^2}\ln^2{u\over s}
+{1\over u}\ln{t\over s}+{1\over t}\ln{u\over s}
\eq
\bq
X^3_2=-~{t\over4u^3}\ln^2{t\over s}+{u\over4t^3}\ln^2{u\over s}
-{s+3t\over 4tu^2}\ln{t\over s}+{s+3u\over 4ut^2}\ln{u\over s}
\eq
\bq
X^3_3=-~{t\over2u^2}\ln^2{t\over s}+{u\over2t^2}\ln^2{u\over s}
-{1\over u}\ln{t\over s}+{1\over t}\ln{u\over s}
\eq

\bq
X^4_1={s\over 2u^2}\ln^2{t\over s}+{s\over2t^2}\ln^2{u\over s}
-{1\over u}\ln{t\over s}-{1\over t}\ln{u\over s}
\eq
\bq
X^4_2={t-s\over4tu^2}\ln{t\over s}-{s\over4u^3}\ln^2{t\over s}
-{u-s\over4ut^2}\ln{u\over s}+{s\over4t^3}\ln^2{u\over s}
\eq
\bq
X^4_3=-~{s\over2u^2}\ln^2{t\over s}+{s\over2t^2}\ln^2{u\over s}
+{1\over u}\ln{t\over s}-{1\over t}\ln{u\over s}
\eq

\bq
X^5_1={2s\over tu}\ln^2{t\over u}-~{6\over u}\ln{u\over M^2}
-~{6\over t}\ln{t\over M^2}+{2t+u\over tu}\ln^2{t\over M^2}
+{2u+t\over tu}\ln^2{u\over M^2}
\eq
\bq
X^5_2={3\over ut}\ln{t\over u}
\eq
\bq
X^5_3={2(u-t)\over tu}\ln^2{t\over u}
+2({1\over t}\ln{t\over M^2}-{1\over u}\ln{u\over M^2})
+2({1\over u}\ln^2{t\over M^2}-{1\over t}\ln^2{u\over M^2})
\eq

\bq
X^6_1=X^6_3=0
\eq
\bq
X^6_2={1\over2tu}\ln{u\over t}
\eq

{\bf \underline{Leading $\ln s$ and $\ln^2s$ terms}}\\

Keeping in the above expressions only the terms proportional to
$\ln(s/ M^2)$ and 
$\ln^2(s/ M^2)$, one obtains:

\bq
B^1_{\lambda,-\lambda}=2[{\lambda+(2\tau)\cos\theta\over \sin\theta}]
[\ln^2{s\over M^2}-4\ln{s\over M^2}]
\eq
\bq
B^1_{\lambda,\lambda}=8[{(2\tau)\cos\theta\over \sin\theta}]
\ln{s\over M^2}
\eq
\bqa
B^2_{\lambda,-\lambda}&=&[{\lambda+(2\tau)\cos\theta\over \sin\theta}]
\{4[\ln^2{s\over M^2}-\ln{s\over M^2}]\nonumber\\
&&+2\sin^2\theta[{1\over1-\cos\theta}\ln{1-\cos\theta\over2}
+{1\over1+\cos\theta}\ln{1+\cos\theta\over2}
-1]\}\ln{s\over M^2}
\eqa

\bqa
B^2_{\lambda,\lambda}&=&-~{2(2\tau)\cos\theta \over \sin\theta}
\ln^2{s\over M^2}-4(2\tau)\sin\theta \ln{1-\cos\theta\over1+\cos\theta}
\ln{s\over M^2}\nonumber\\
&&+2(2\tau)\cos\theta \sin\theta
[1-{1\over1-\cos\theta}\ln{1-\cos\theta\over2}
-{1\over1+\cos\theta}\ln{1+\cos\theta\over2}
]\ln{s\over M^2}
\eqa

\bqa
B^5_{\lambda,-\lambda}&=&[{\lambda+(2\tau)\cos\theta\over \sin\theta}]
\{6(-\ln^2{s\over M^2}+2\ln{s\over M^2})\nonumber\\
&&-2[
(3-\cos\theta)\ln{1-\cos\theta\over2}
+(3+\cos\theta)\ln{1+\cos\theta\over2}]\ln{s\over M^2}\}
\eqa

\bqa
B^5_{\lambda,\lambda}&=&2[{(2\tau)\cos\theta\over \sin\theta}]
\{\ln^2{s\over M^2}-4\ln{s\over M^2}\nonumber\\
&&
+(3-\cos\theta)\ln{1-\cos\theta\over2}
+(3+\cos\theta)\ln{1+\cos\theta\over2}]\ln{s\over M^2}\}\nonumber\\
&&
+4(2\tau)\sin\theta[{1\over1+\cos\theta}\ln{1-\cos\theta\over2}
-{1\over1-\cos\theta}\ln{1+\cos\theta\over2}]\ln{s\over M^2}
\eqa

Using these simple expressions in 
eqs.(\ref{gZm},\ref{gZp},\ref{Wm},\ref{Wp},\ref{Hm},\ref{Hp},\ref{Sm},\ref{Sp},\ref{SYm},\ref{SYp}),
one obtains the leading terms of the helicity amplitudes given
in eqs.(\ref{gZmlt},\ref{gZplt},\ref{Wmlt},
\ref{Wplt},\ref{Hmlt},\ref{Hplt},\ref{Smlt},\ref{Splt}
,\ref{SYmlt},\ref{SYplt}).

\newpage

\section{The polarized $\gamma\gamma\to 
\lowercase{f\bar f}$ cross section}

In the high energy limit, with real helicity amplitudes, the general
expression of the polarized $\gamma\gamma$ cross section \cite{ggsig}
is:

\bqa
{d\sigma\over d\tau d\cos\theta}&=&{d \bar L_{\gamma\gamma}\over
d\tau} \Bigg \{
{d\bar{\sigma}_0\over d\cos\theta}
+\langle \xi_2 
\rangle{d\bar{\sigma}_{2}\over d\cos\theta}
+\langle \xi^\prime _2 
\rangle{d\bar{\sigma^\prime}_{2}\over d\cos\theta}
+\langle \xi_2 \xi^\prime_2
\rangle{d\bar{\sigma}_{22}\over d\cos\theta}\nonumber\\
&&
+\langle\xi_3\rangle \cos2\phi \,
{d\bar{\sigma}_{3}\over d\cos\theta}
+\langle\xi_3^ \prime\rangle\cos2\phi^\prime
{d\bar\sigma_3^\prime\over d\cos\theta}
\nonumber\\
&&+\langle\xi_3 \xi_3^\prime\rangle
\left[{d\bar{\sigma}_{33}\over d\cos\theta}
\cos2(\phi+\phi^\prime)
+{d\bar{\sigma}^\prime_{33}\over
d\cos\theta^*}\cos2(\phi- \phi^\prime)\right ]\nonumber\\
&&+ \langle\xi_2 \xi_3^\prime\rangle\cos2 \phi^\prime
{d\bar{\sigma}_{23}\over d\cos\theta}-
\langle\xi_3 \xi^\prime_2\rangle\cos2\phi\,
{d\bar{\sigma^\prime}_{23}\over d\cos\theta} \Bigg \} \ \ ,
\label{sigpol}
\eqa
In (\ref{sigpol}), $\tau=s /s_{ee}$, where
$s\equiv s_{\gamma \gamma}$, while
$ d \bar L_{\gamma\gamma}/d\tau$ describes the photon-photon
luminosity per unit $e^-e^+$ flux
\cite{laser}. The Stokes parameters
$(\xi_2, \xi^\prime_2)$, $(\xi_3,~ \xi^\prime_3)$
and $(\phi,~ \phi^\prime)$ describe respectively
 the average helicities, transverse
polarizations and azimuthal angles of the two
 backscattered photons. Typical values for these various quantities
are given in ref.\cite{ggsig}. In
(\ref{sigpol}) there appear the following quantities

\bqa
{d\bar \sigma_0\over d\cos\theta}&=&
\left ({N_f\over128\pi s}\right )
\sum_{\lambda_3\lambda_4} [|F_{++\lambda_3\lambda_4}|^2
+|F_{--\lambda_3\lambda_4}|^2+|F_{+-\lambda_3\lambda_4}|^2
+|F_{-+\lambda_3\lambda_4}|^2] ~ ,  \label{sig0} \\
{d\bar{\sigma}_{2}\over d\cos\theta} &=&
\left ({N_f\over128\pi s}\right )\sum_{\lambda_3\lambda_4}
[|F_{++\lambda_3\lambda_4}|^2-[|F_{--\lambda_3\lambda_4}|^2
+|F_{+-\lambda_3\lambda_4}|^2-|F_{-+\lambda_3\lambda_4}|^2]
  \ , \label{sig2} \\
{d\bar{\sigma^\prime}_{2}\over d\cos\theta} &=&
\left ({N_f\over128\pi s}\right )\sum_{\lambda_3\lambda_4}
[|F_{++\lambda_3\lambda_4}|^2-[|F_{--\lambda_3\lambda_4}|^2
-|F_{+-\lambda_3\lambda_4}|^2+|F_{-+\lambda_3\lambda_4}|^2]
  \ , \label{sig2p} \\
{d\bar{\sigma}_{22}\over d\cos\theta} &=&
\left ({N_f\over128\pi s}\right )\sum_{\lambda_3\lambda_4}
[|F_{++\lambda_3\lambda_4}|^2+[|F_{--\lambda_3\lambda_4}|^2
-|F_{+-\lambda_3\lambda_4}|^2-|F_{-+\lambda_3\lambda_4}|^2]
  \ , \label{sig22} \\
{d\bar{\sigma}_{3} \over d\cos\theta} &=&
\left ({-N_f\over64\pi s}\right ) \sum_{\lambda_3\lambda_4}
[F_{++\lambda_3\lambda_4}F_{-+\lambda_3\lambda_4}]
+[F_{--\lambda_3\lambda_4}F_{+-\lambda_3\lambda_4}]  \ ,
\label{sig3p} \\
{d\bar{\sigma^\prime}_{3}\over d\cos\theta} &=&
\left ({-N_f\over64\pi s}\right ) \sum_{\lambda_3\lambda_4}
[F_{++\lambda_3\lambda_4}F_{+-\lambda_3\lambda_4}] 
+[F_{--\lambda_3\lambda_4}F_{-+\lambda_3\lambda_4}]  \ ,
\label{sig3} \\
{d\bar \sigma_{33} \over d\cos\theta}& = &
\left ({N_f\over64\pi s}\right ) \sum_{\lambda_3\lambda_4}
[F_{+-\lambda_3\lambda_4}F_{-+\lambda_3\lambda_4}] \ ,
\label{sig33} \\
{d\bar{\sigma^\prime}_{33}\over d\cos\theta} &=&
\left ({N_f\over64\pi s}\right ) \sum_{\lambda_3\lambda_4}
[F_{++\lambda_3\lambda_4}F_{--\lambda_3\lambda_4}] \  ,
\label{sig33prime} \\
{d\bar{\sigma}_{23}\over d\cos\theta}& = &
\left ({-N_f\over 64\pi s}\right ) \sum_{\lambda_3\lambda_4}
[F_{++\lambda_3\lambda_4}F_{+-\lambda_3\lambda_4}]
-[F_{--\lambda_3\lambda_4}F_{-+\lambda_3\lambda_4}] \ ,
\label{sig23}\\
{d\bar{\sigma^\prime}_{23} \over d\cos\theta}& = &
\left ({-N_f\over 64 \pi s}\right ) \sum_{\lambda_3\lambda_4}
[F_{++\lambda_3\lambda_4}F_{-+\lambda_3\lambda_4}]
-[F_{--\lambda_3\lambda_4}F_{+-\lambda_3\lambda_4}] \ ,
\label{sig23p}
\eqa
\noindent
where $N_f$ is the colour factor ($3$ when f is a quark and $1$ 
when it is a lepton).\\

Using the fact that at high energy the only non vanishing fermion
helicities are $\lambda_3=-\lambda_4\equiv\tau$,
as well as the relations due to Bose symmetry and CP-conservation,

\bq
F_{+,-,\tau,-\tau}(s,cos\theta)=-~F_{-+,\tau,-\tau}(s,-cos\theta)
\eq
\bq
F_{++,\tau,-\tau}(s,cos\theta)=
F_{--,\tau,-\tau}(s,cos\theta)=-~F_{++,\tau,-\tau}(s,-cos\theta)
=-~F_{--,\tau,-\tau}(s,-cos\theta)
\eq
one sees that\\
$${d\bar \sigma_0\over d\cos\theta},~~
{d\bar \sigma_3\over d\cos\theta}\equiv
{d\bar \sigma_3^\prime\over d\cos\theta^*},~~
{d\bar \sigma_{22}\over d\cos\theta},~~
{d\bar \sigma_{33}\over d\cos\theta},~~
{d\bar \sigma_{33}^\prime\over d\cos\theta}$$ 
are \underline{$\cos\theta$-symmetric},\\
and that\\ 
$${d\bar \sigma_2\over d\cos\theta}\equiv
-~{d\bar \sigma^\prime_2\over d\cos\theta},~~
{d\bar \sigma_{23}\over d\cos\theta}\equiv
-~{d\bar \sigma^\prime_{23}\over d\cos\theta}$$
are \underline{$\cos\theta$-antisymmetric}.\\

The \underline{Born amplitudes} are such that
\bq
F^{Born}_{\lambda,\lambda,\tau,-\tau}(s,cos\theta)=0
\eq
\bq
F^{Born}_{\lambda,-\lambda,\tau,-\tau}(s,cos\theta)
=-F^{Born}_{-\lambda,\lambda,-\tau,\tau}(s,cos\theta)
\eq
leading to the only non vanishing Born contributions
\bq
{d\bar \sigma^{Born}_0\over d\cos\theta}\equiv
-{d\bar \sigma^{Born}_{22}\over d\cos\theta},
~~~~~~~{d\bar \sigma^{Born}_{33}\over d\cos\theta}
\eq
\underline{At first order ($\alpha^3$) in the electroweak corrections}
(i.e. neglecting the terms quadratic in 
$F_{\lambda,\lambda,\tau,-\tau}$), one 
has the additional properties:
\bq
{d\bar \sigma_0\over d\cos\theta}=
-{d\bar \sigma_{22}\over d\cos\theta},
~~~~~~~{d\bar \sigma^\prime_{33}\over d\cos\theta}=0
\eq

So that only \underline{five} observables remain:\\

~~~--- The 3 symmetric ones:\\
$${d\bar \sigma_0\over d\cos\theta}\equiv
-{d\bar \sigma_{22}\over d\cos\theta},~~~
{d\bar \sigma_3\over d\cos\theta}\equiv
{d\bar \sigma^\prime_{3}\over d\cos\theta},~~~
{d\bar \sigma_{33}\over d\cos\theta}$$\\

~~~--- The 2 antisymmetric ones\\
$${d\bar \sigma_2\over d\cos\theta}\equiv
-~{d\bar \sigma^\prime_{2}\over d\cos\theta},~~~
{d\bar \sigma_{23}\over d\cos\theta}\equiv
-~{d\bar \sigma^\prime_{23}\over d\cos\theta}\ \ \ .$$

\begin{figure}[p]

$$\epsfig{file=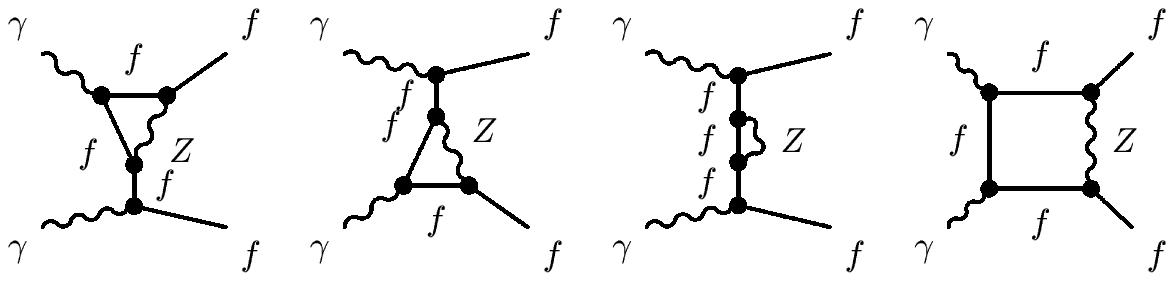}$$
\centerline{(a)}

$$\epsfig{file=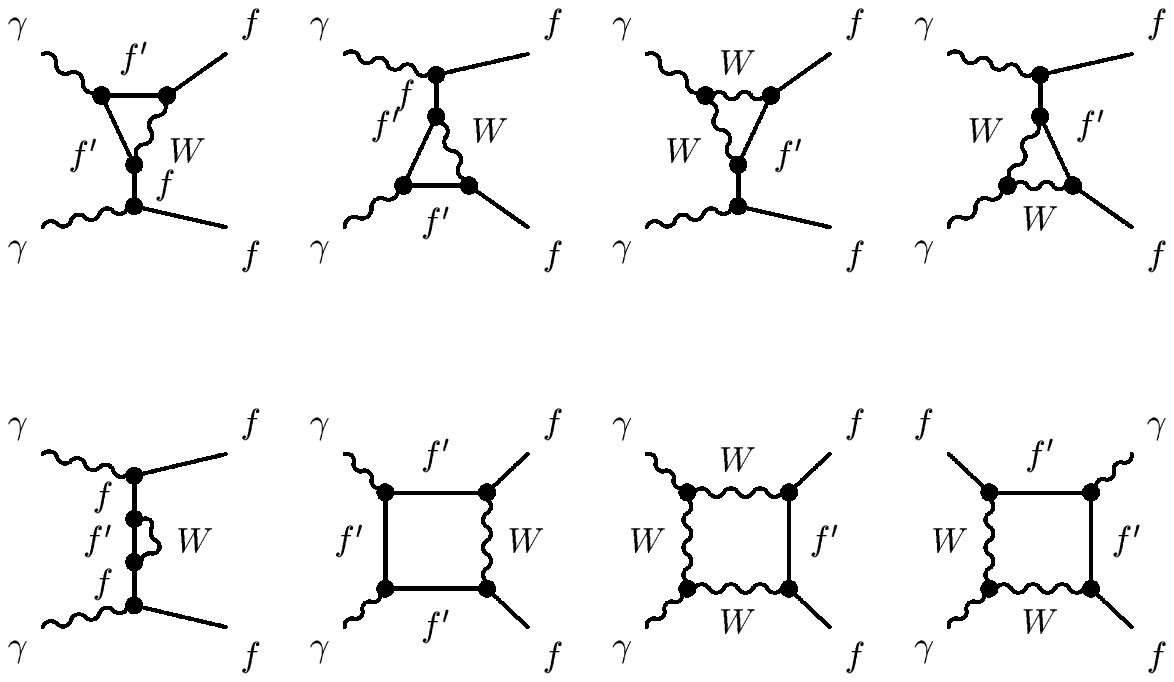}$$
\centerline{(b)}

$$\epsfig{file=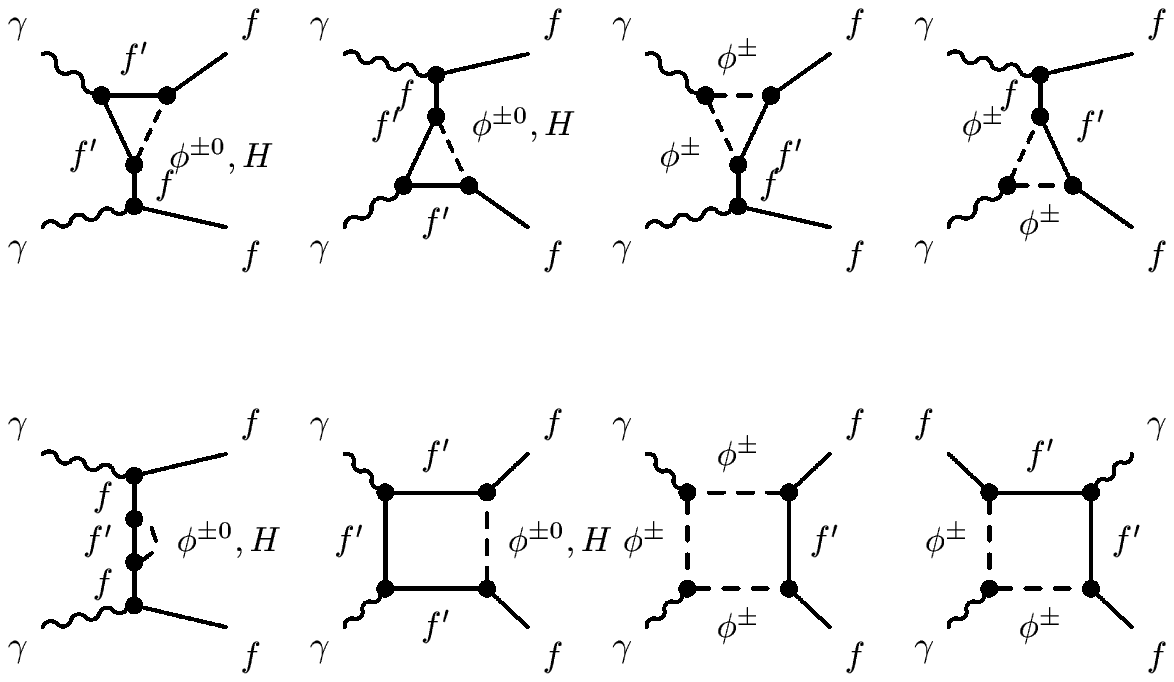}$$
\centerline{(c)}
\vspace{0.2cm}\null


\caption[1]{SM diagrams contributing in the asymptotic regime of
$\gamma\gamma\to f\bar f$, $Z$ sector (a), $W$ sector (b), Higgs
sector (c).}
\label{diagSM}

\end{figure}

\begin{figure}[p]
\vspace*{2.cm}
$$\epsfig{file=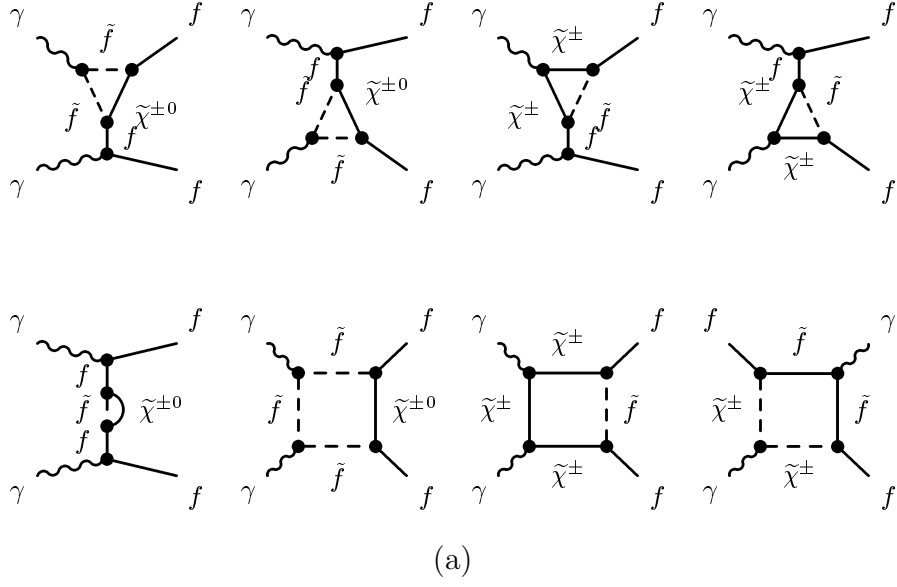}$$
\centerline{(a)}
\vspace{0.5cm}\null

$$\epsfig{file=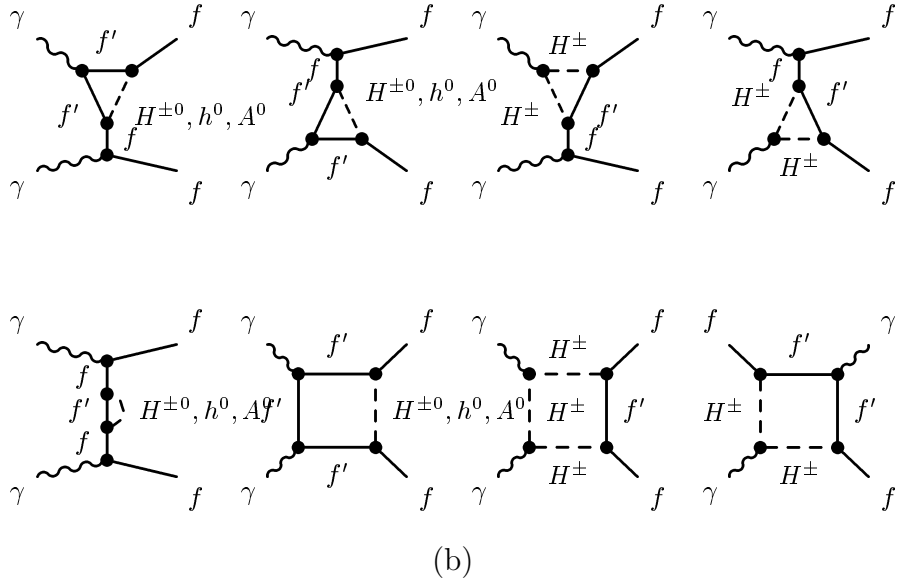}$$
\centerline{(b)}
\vspace{0.5cm}\null


\caption[2]{SUSY additional diagrams contributing 
in the asymptotic regime of
$\gamma\gamma\to f\bar f$, Chargino and neutralino sector (a), 
SUSY Higgs sector (b).}
\label{diagSUSY}
\end{figure}


\begin{figure}[p]
$$\epsfig{file=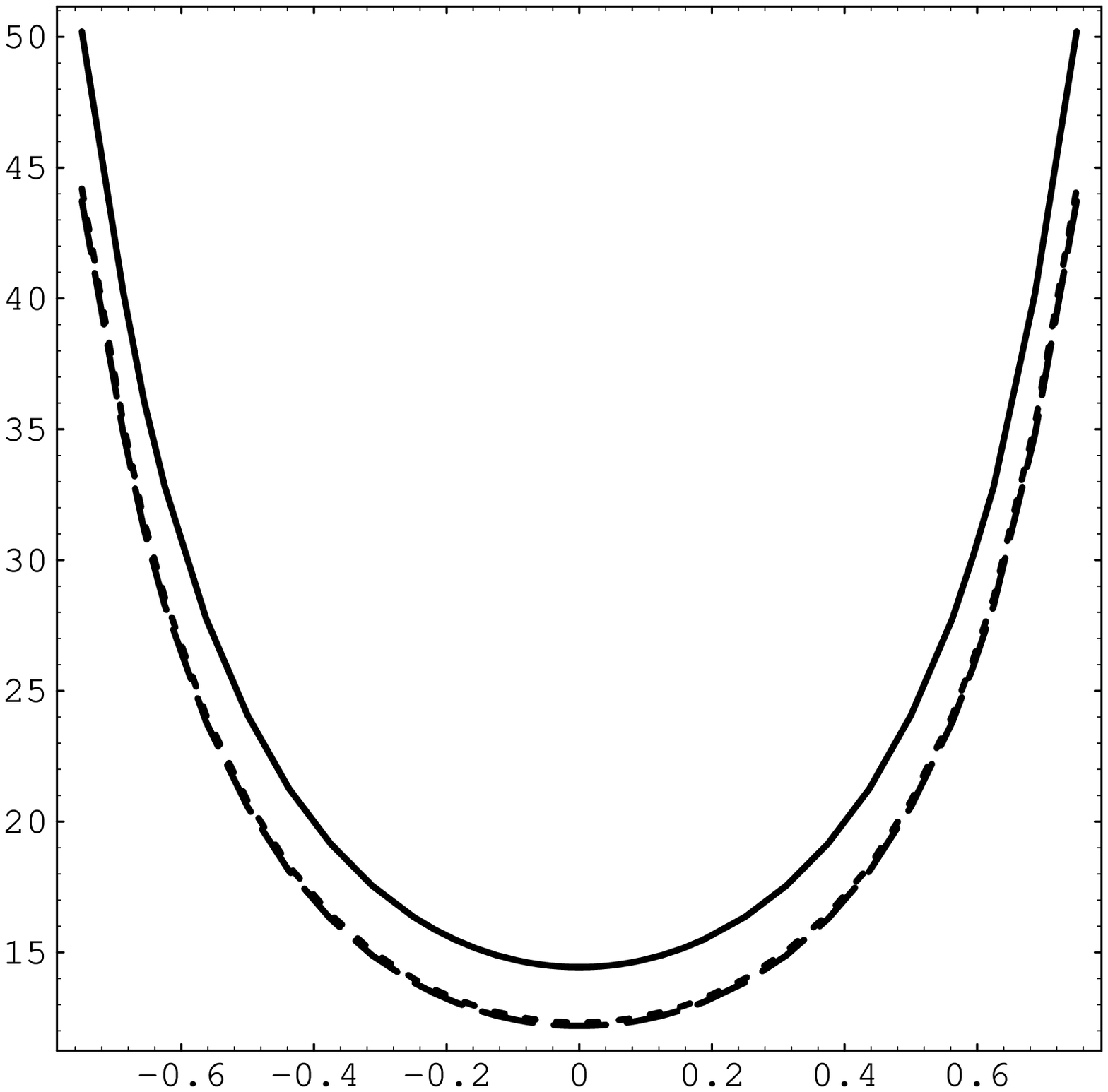,height=9.cm}\hspace{3.5cm}
\epsfig{file=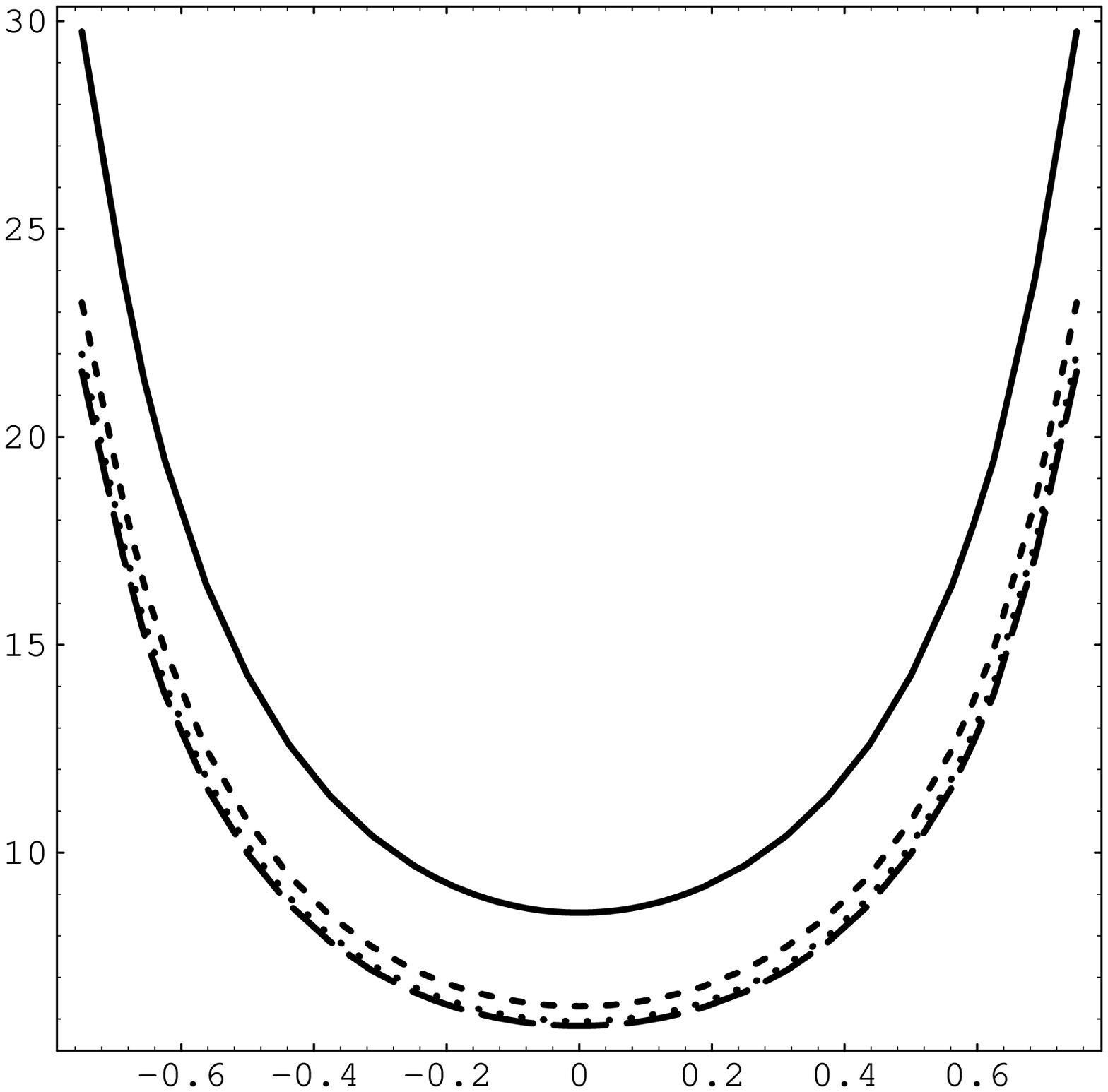,height=9.cm}$$
\vspace{-6cm}\null

\hspace*{-1.5cm}
 \begin{sideways}
 {\small $ d\sigma_0/d\cos\theta$ }
 \end{sideways}
\hspace{9.5cm}
 \begin{sideways}
 {\small $ d\sigma_0/d\cos\theta$ }
 \end{sideways}
\vspace{2.cm}\null

\centerline{\small $\cos\theta$ \hspace{9.5cm} \small $\cos\theta$ }
\vspace{0.2cm}

\centerline{\small (a) \hspace{10cm} \small (b)}

$$\epsfig{file=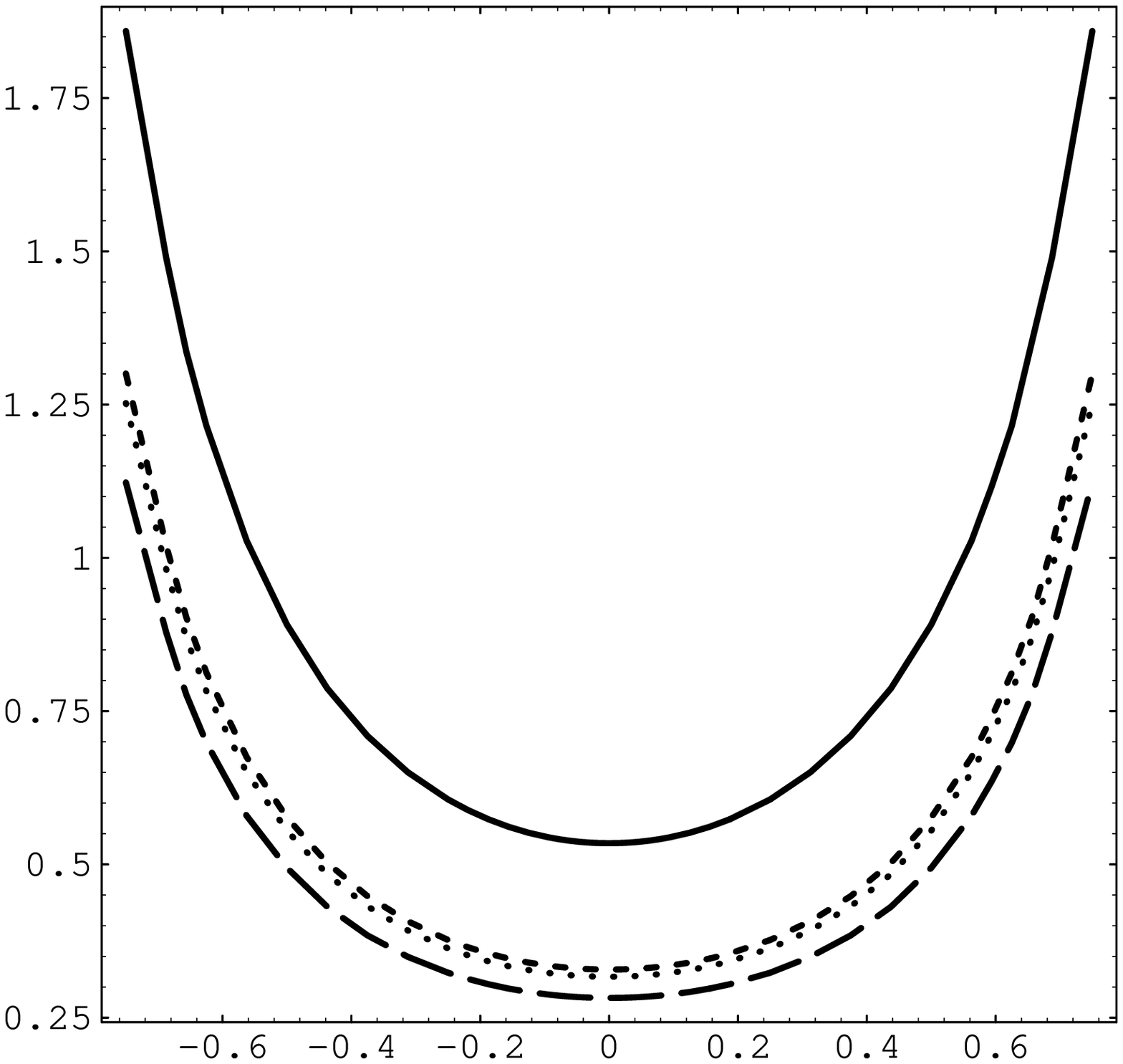,height=9.cm}$$

\vspace{-6cm}\null

\hspace*{3.8cm}
 \begin{sideways}
 {\small $ d\sigma_0/d\cos\theta$ }
 \end{sideways}

\vspace{1.6cm}\null

\centerline{\small $\cos\theta$}
\vspace{0.2cm}

\centerline{\small (c)}
\vspace{0.2cm}\null


\caption[3] { Angular distribution of the unpolarized
$\gamma\gamma\to f\bar f$ cross section at 3 TeV; $l^+l^-$ (a), $t\bar
t$ (b), $b\bar b$ (c); Born (solid), total SM (small dashed),
total MSSM($\tan\beta=4$) (dotted), total MSSM($\tan\beta=40$) (large
dashed).}
\label{ang0}
\end{figure}


\begin{figure}[p]
$$\epsfig{file=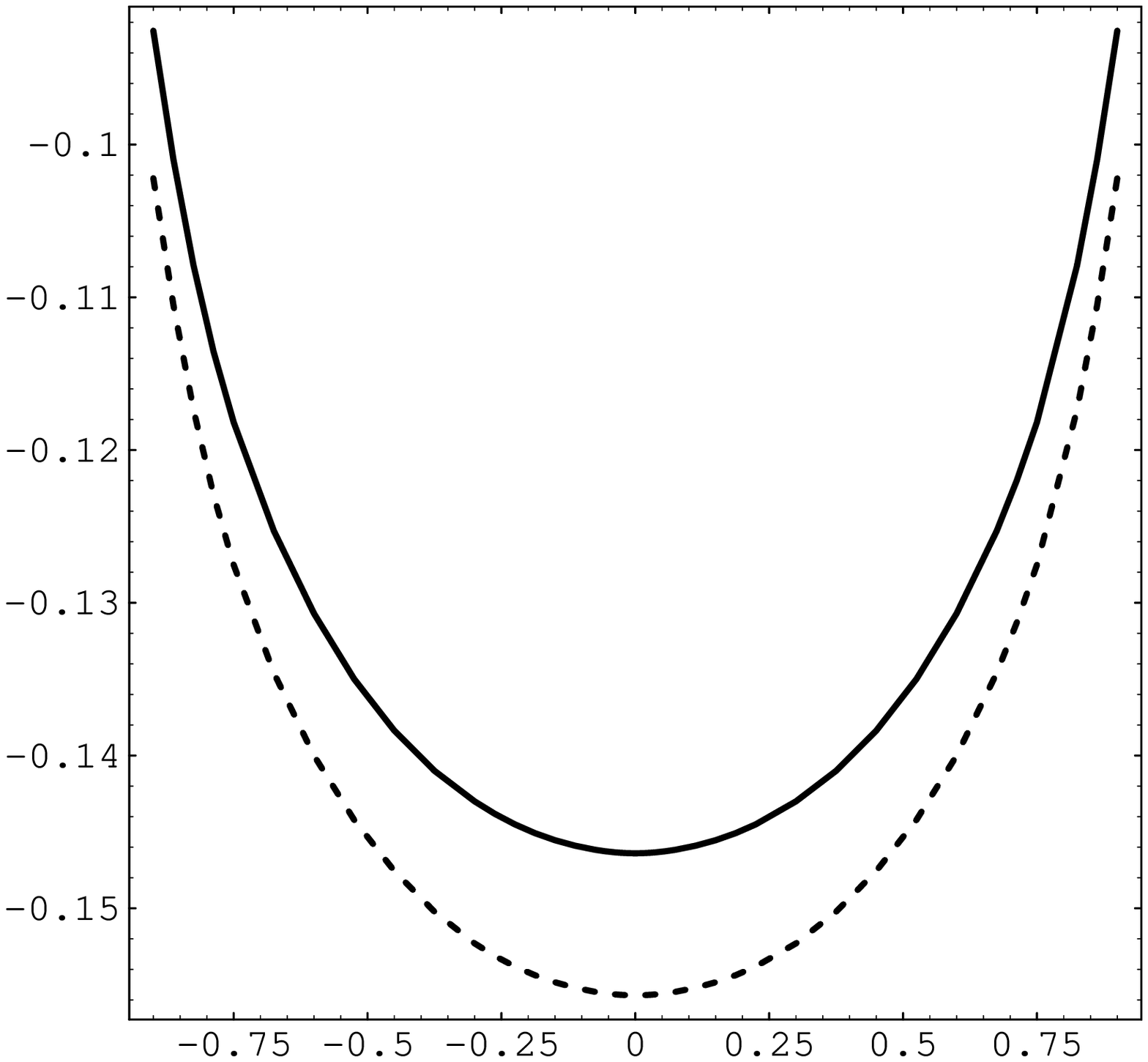,height=9.cm}\hspace{3.5cm}
\epsfig{file=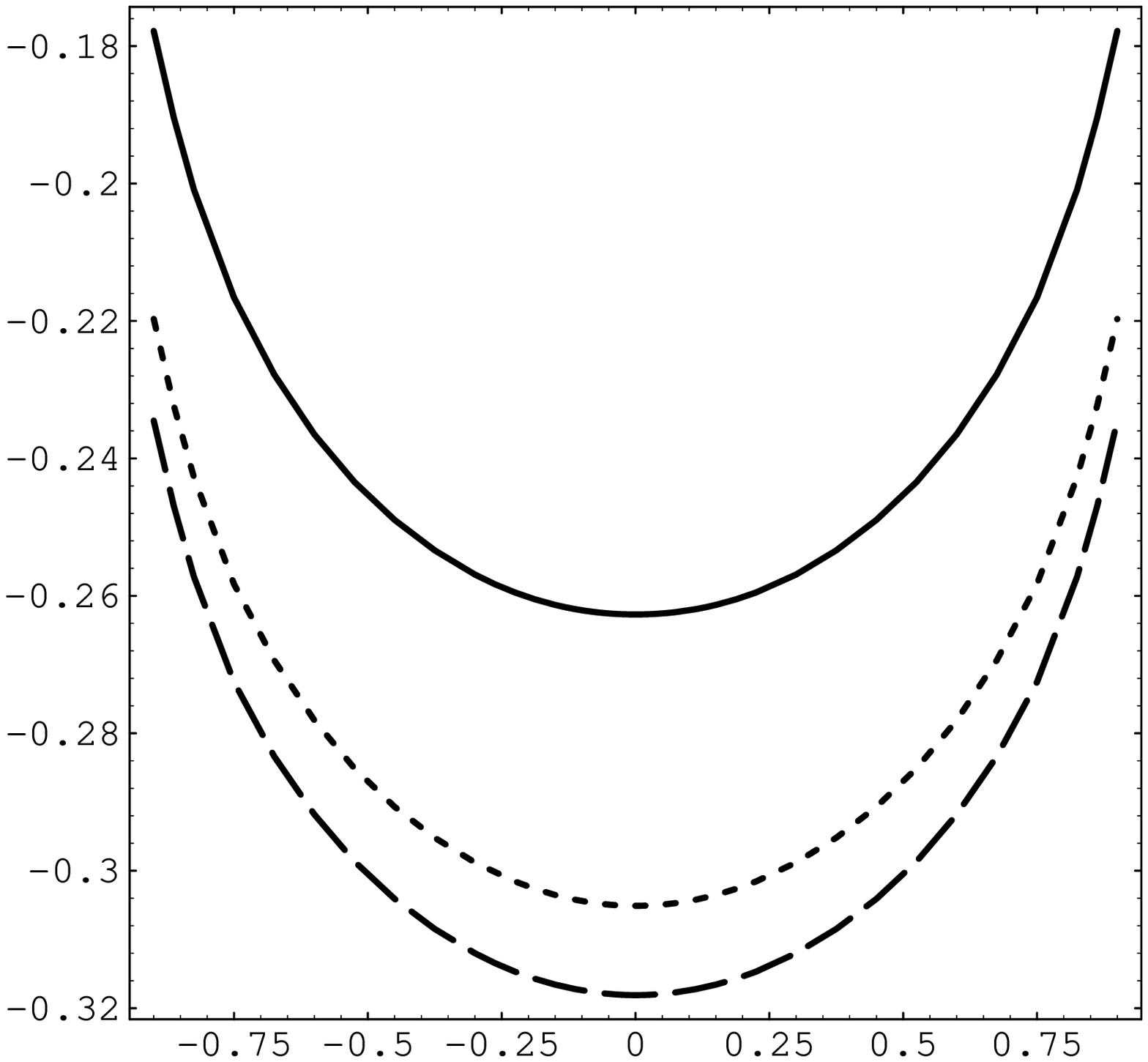,height=9.cm}$$
\vspace{-6cm}\null

\hspace*{-1.5cm}
 \begin{sideways}
 {\small $\Delta( d\sigma_0/d\cos\theta)$ }
 \end{sideways}
\hspace{9.5cm}
 \begin{sideways}
 {\small $\Delta( d\sigma_0/d\cos\theta)$ }
 \end{sideways}
\vspace{1.2cm}\null

\centerline{\small $\cos\theta$ \hspace{9.5cm} \small $\cos\theta$ }
\vspace{0.1cm}\null

\centerline{\small (a) \hspace{10cm} \small (b)}

$$\epsfig{file=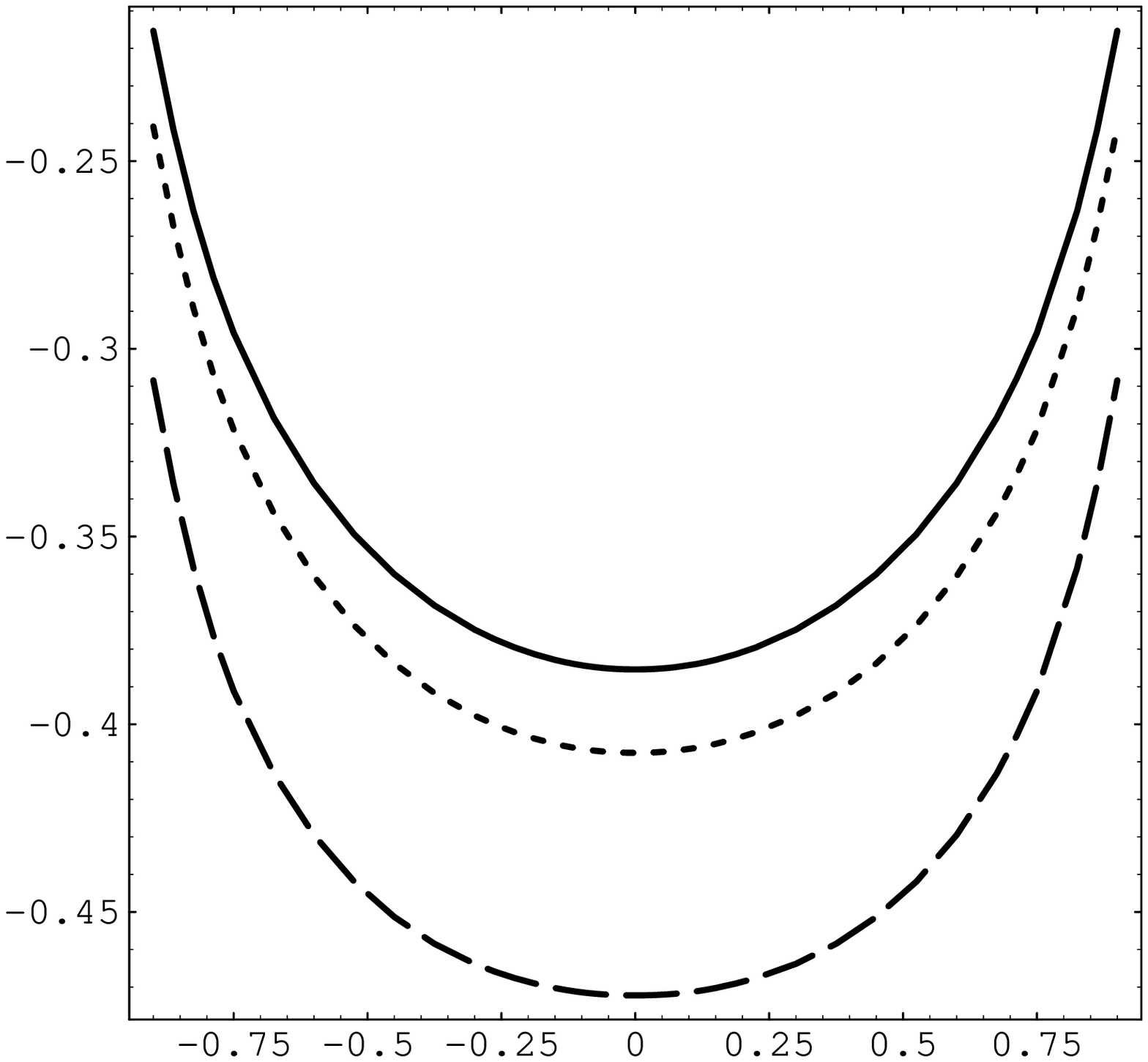,height=9.cm}$$

\vspace{-6.2cm}\null

\hspace*{3.8cm}
 \begin{sideways}
 {\small $\Delta( d\sigma_0/d\cos\theta)$ }
 \end{sideways}

\vspace{1.2cm}\null

\centerline{\small $\cos\theta$}
\vspace{0.1cm}\null

\centerline{\small (c)}
\vspace{0.2cm}\null

\caption[4]{Angular distribution of the relative departure
from the unpolarized Born
$\gamma\gamma\to f\bar f$ cross section at 3 TeV due to electroweak
radiative corrections; $l^+l^-$ (a), $t\bar
t$ (b), $b\bar b$ (c); total SM (solid),
total MSSM($\tan\beta=4$) (small dashed), 
total MSSM($\tan\beta=40$) (large dashed).}
\label{relang0}
\end{figure}


\begin{figure}[p]

$$\epsfig{file=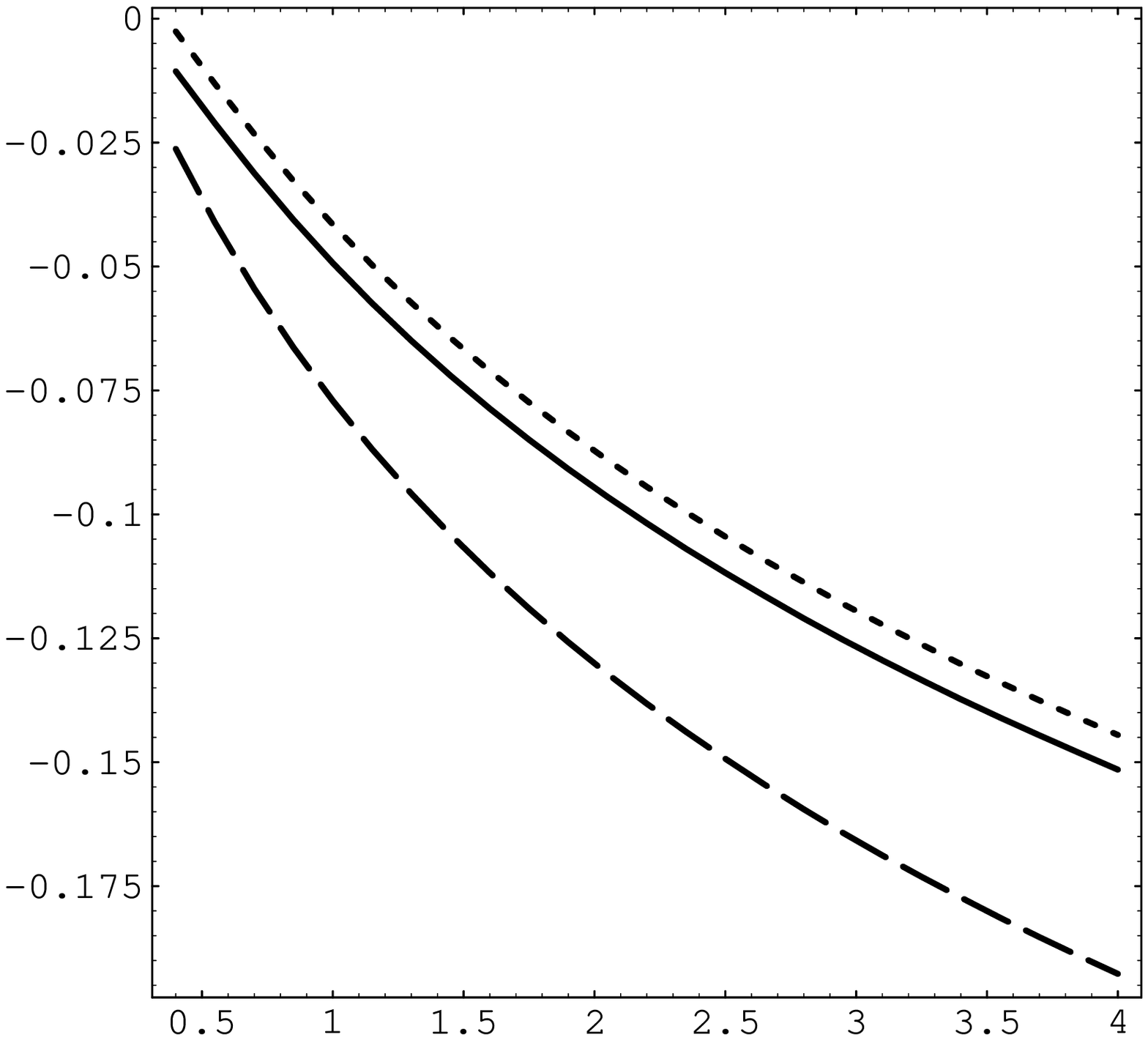,height=9.cm}$$

\vspace{-5.8cm}\null

\hspace*{3.5cm}
 \begin{sideways}
 {\small $ R_0 $ }
 \end{sideways}

\vspace{2.8cm}\null

\centerline{\small $\sqrt{s}\ (TeV)$}
\vspace{0.1cm}\null

\centerline{\small (a)}
\vspace{0.2cm}\null

$$\epsfig{file=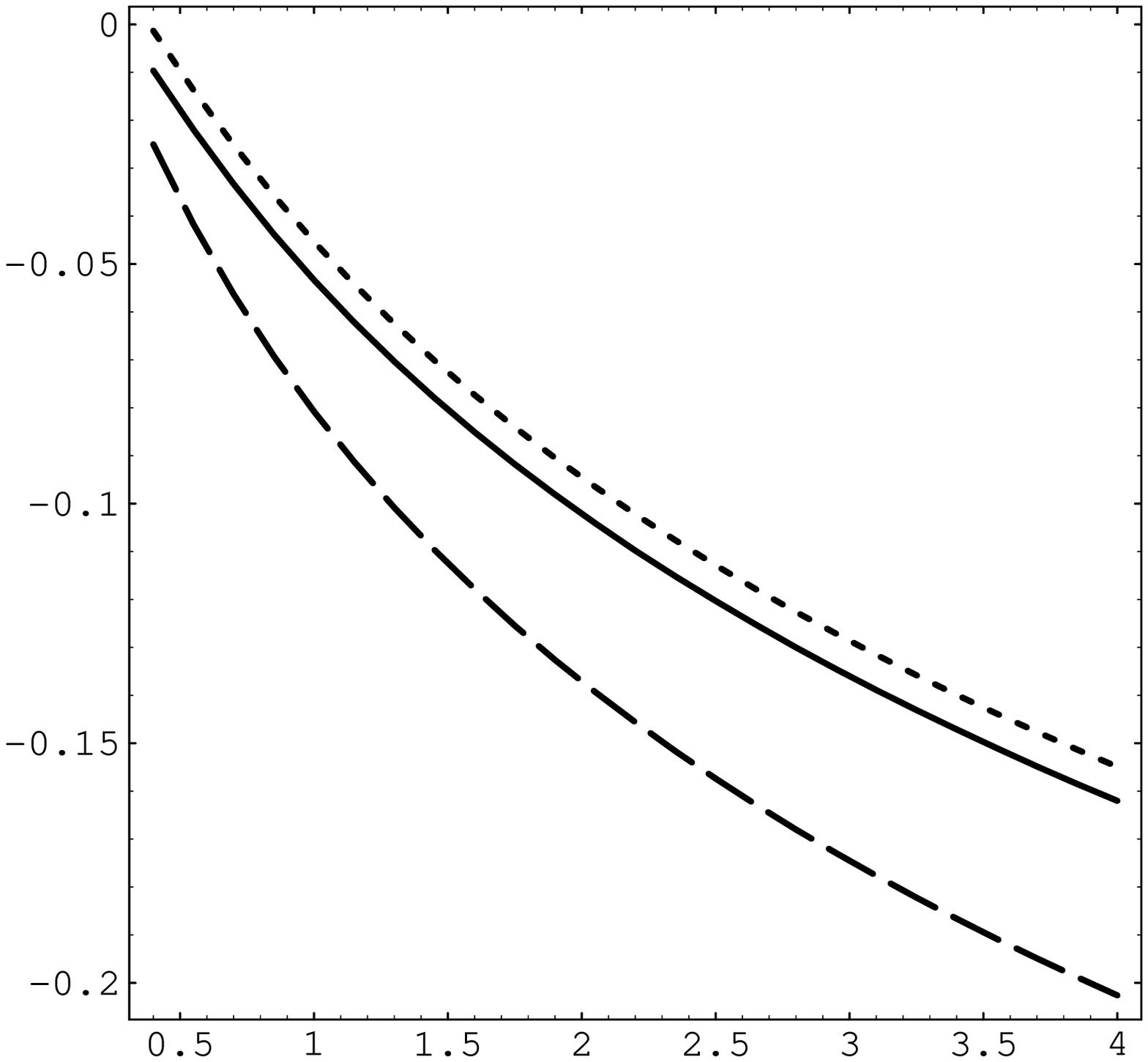,height=9.cm}$$

\vspace{-5.8cm}\null

\hspace*{3.5cm}
 \begin{sideways}
 {\small $ R_0 $ }
 \end{sideways}

\vspace{2.8cm}\null

\centerline{\small $\sqrt{s}\ (TeV)$}
\vspace{0.1cm}\null

\centerline{\small (b)}
\vspace{0.2cm}\null


\caption[5]{The ratio $R_0$ for
$\gamma\gamma\to l^+l^-$ versus the energy; SM (a), MSSM(b);
all logarithmic terms (solid),
leading terms only (small dashed), 
leading angular independent terms only(large dashed).}
\label{R0l}
\end{figure}


\begin{figure}[p]
$$\epsfig{file=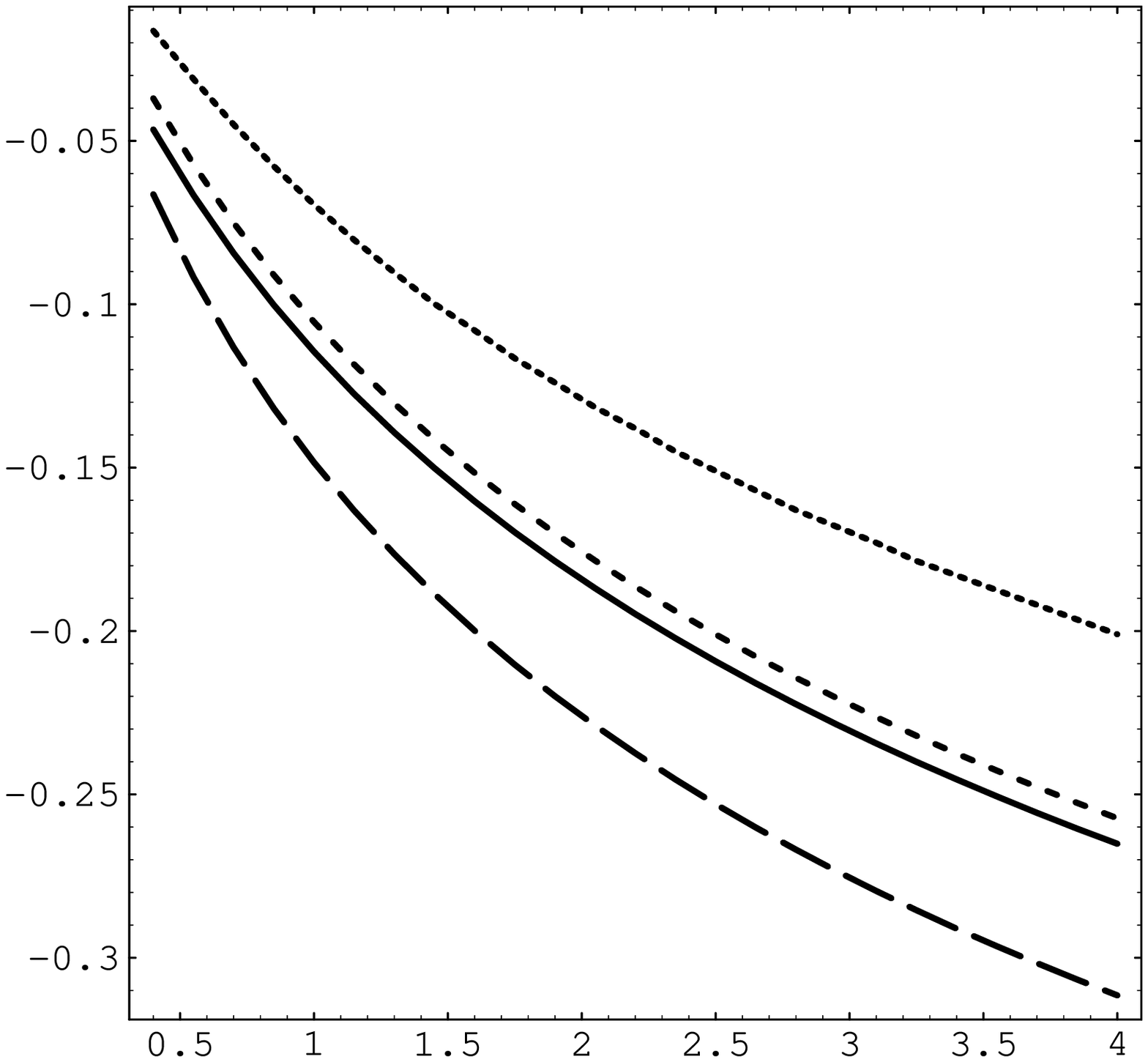,height=9.cm}\hspace{3.5cm}
\epsfig{file=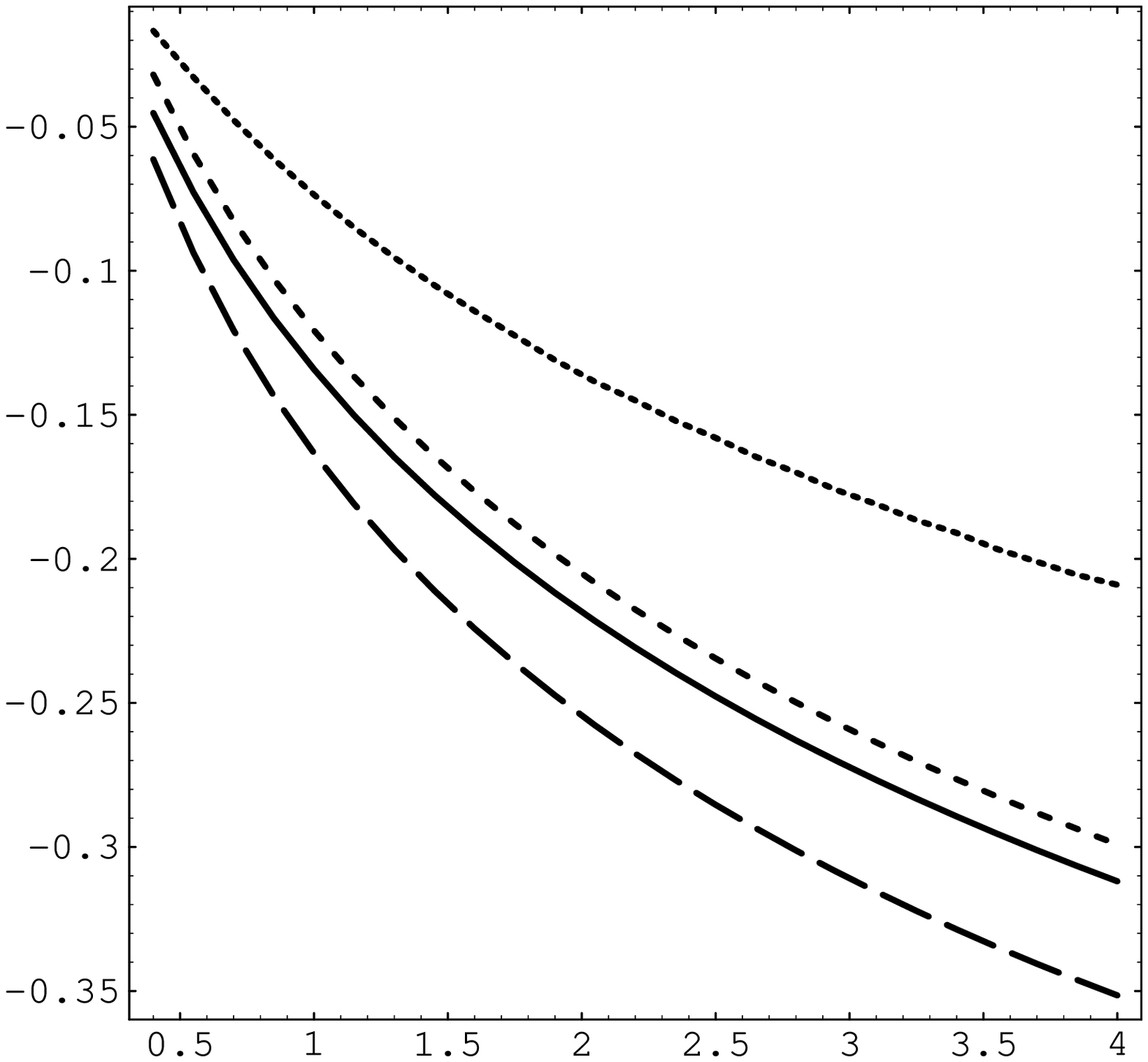,height=9.cm}$$
\vspace{-5.8cm}\null

\hspace*{-1.5cm}
 \begin{sideways}
 {\small $ R_0 $ }
 \end{sideways}
\hspace{9.5cm}
 \begin{sideways}
 {\small $ R_0 $ }
 \end{sideways}
\vspace{3.cm}\null

\centerline{\small $\sqrt{s}\ (TeV)$ \hspace{8.cm} \small $\sqrt{s}\ (TeV)$ }
\vspace{0.1cm}\null

\centerline{\small (a) \hspace{9.5cm} \small (b)}

$$\epsfig{file=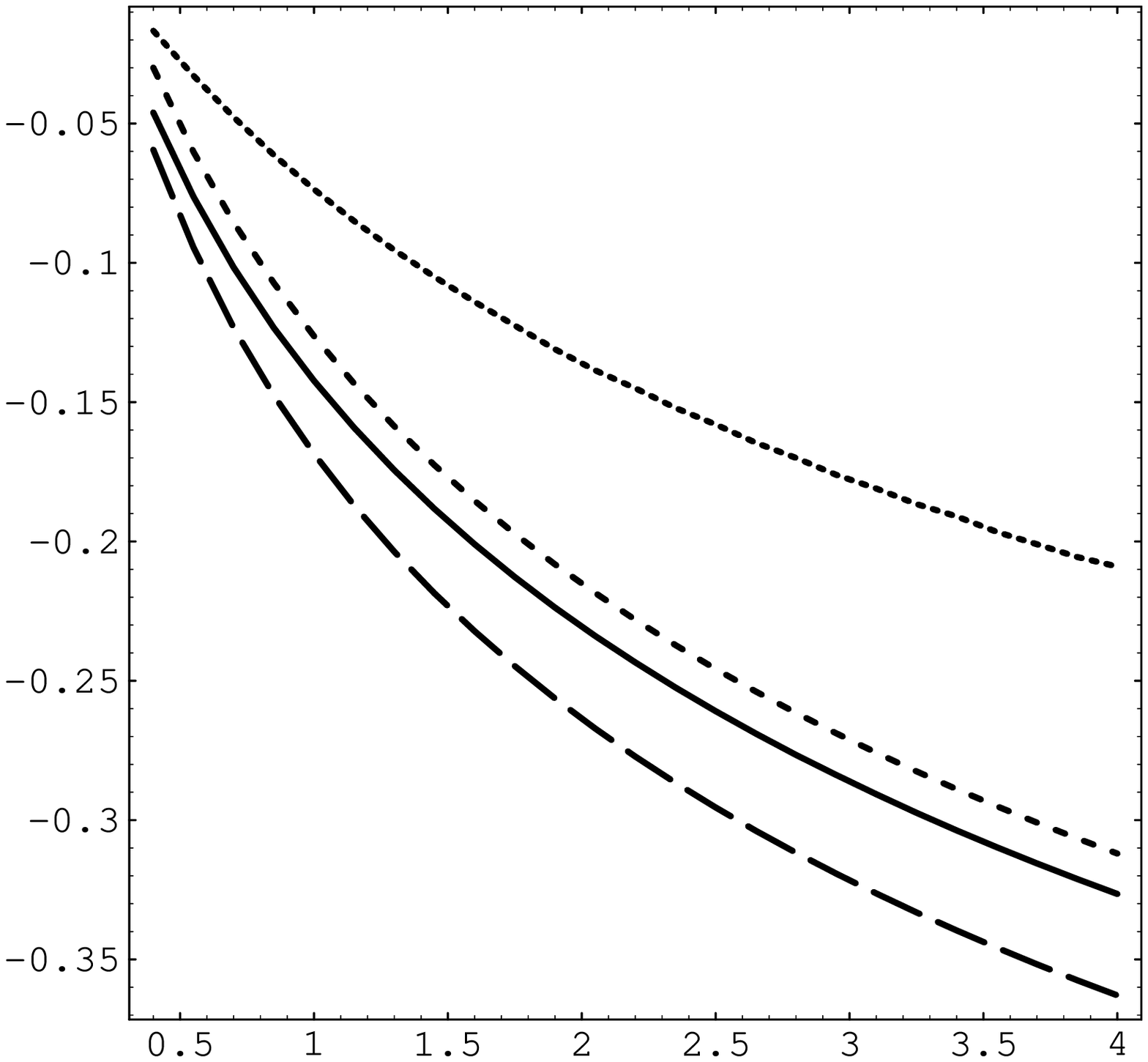,height=9.cm}$$

\vspace{-5.8cm}\null

\hspace*{3.6cm}
 \begin{sideways}
 {\small $ R_0 $ }
 \end{sideways}

\vspace{2.8cm}\null

\centerline{\small $\sqrt{s}\ (TeV)$}
\vspace{0.1cm}\null

\centerline{\small (c)}
\vspace{0.2cm}\null


\caption[6]{The ratio $R_0$ for
$\gamma\gamma\to t\bar t$ versus the energy; SM (a), 
MSSM($\tan\beta=4$) (b); MSSM($\tan\beta=40$) (c);
all logarithmic terms (solid),
leading terms only (small dashed), 
leading angular independent terms only(large dashed); 
all logarithmic without Yukawa terms (very small dashed).}
\label{R0t}
\end{figure}


\begin{figure}[p]
$$\epsfig{file=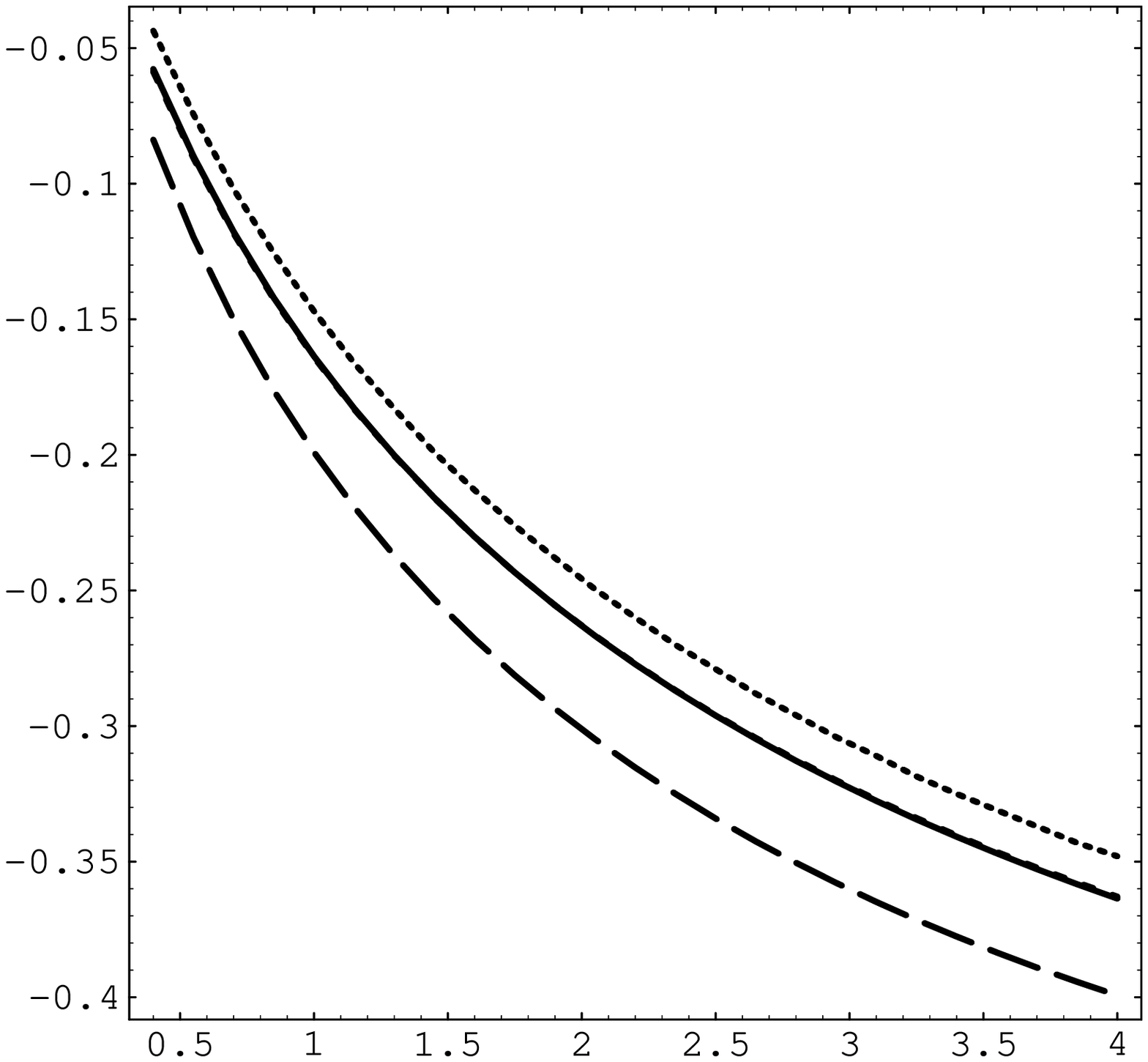,height=9.cm}\hspace{3.5cm}
\epsfig{file=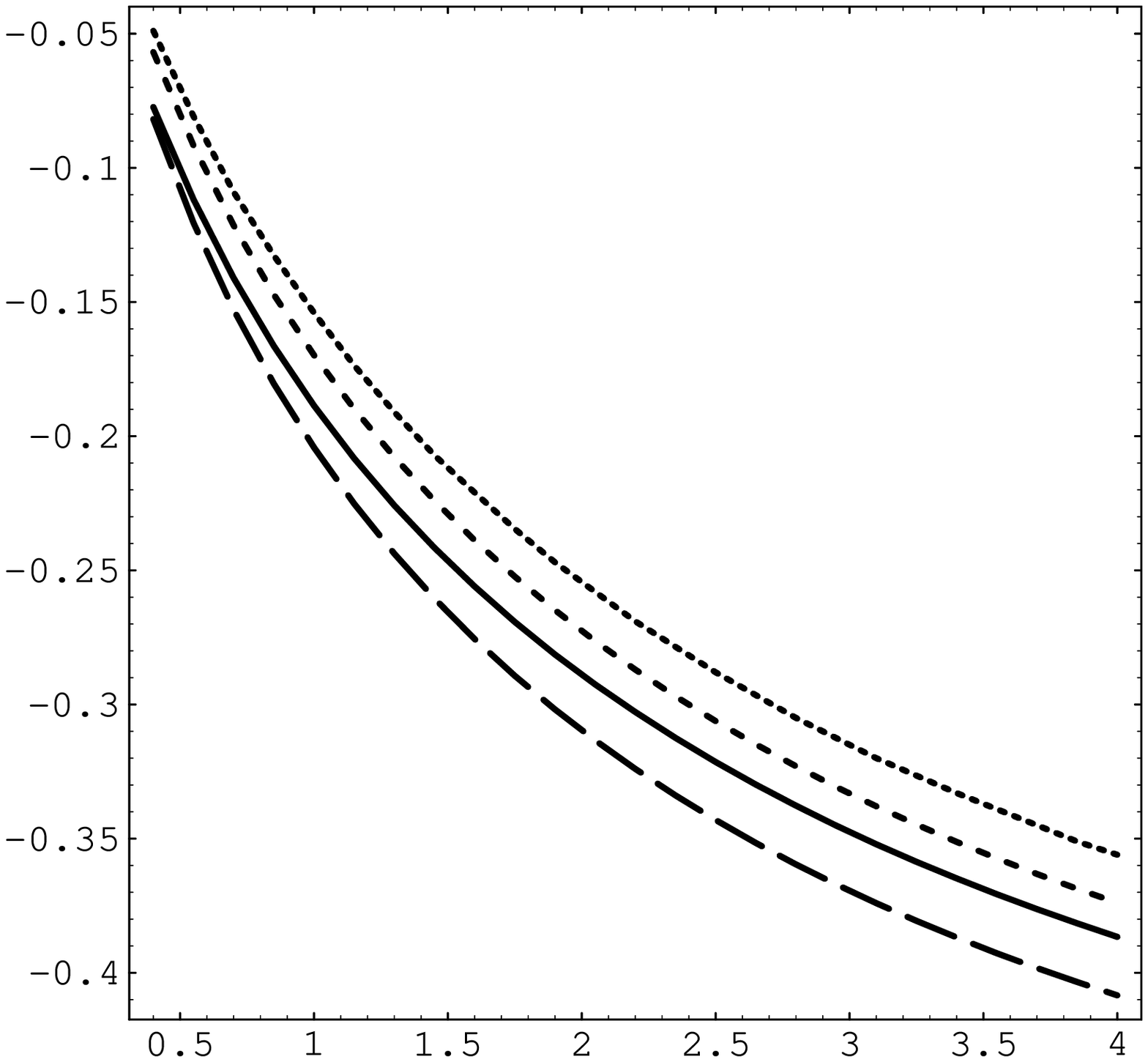,height=9.cm}$$
\vspace{-5.8cm}\null

\hspace*{-1.5cm}
 \begin{sideways}
 {\small $ R_0 $ }
 \end{sideways}
\hspace{9.5cm}
 \begin{sideways}
 {\small $ R_0 $ }
 \end{sideways}
\vspace{3.cm}\null

\centerline{\small $\sqrt{s}\ (TeV)$ \hspace{8.cm} \small $\sqrt{s}\ (TeV)$ }
\vspace{0.1cm}\null

\centerline{\small (a) \hspace{9.5cm} \small (b)}

$$\epsfig{file=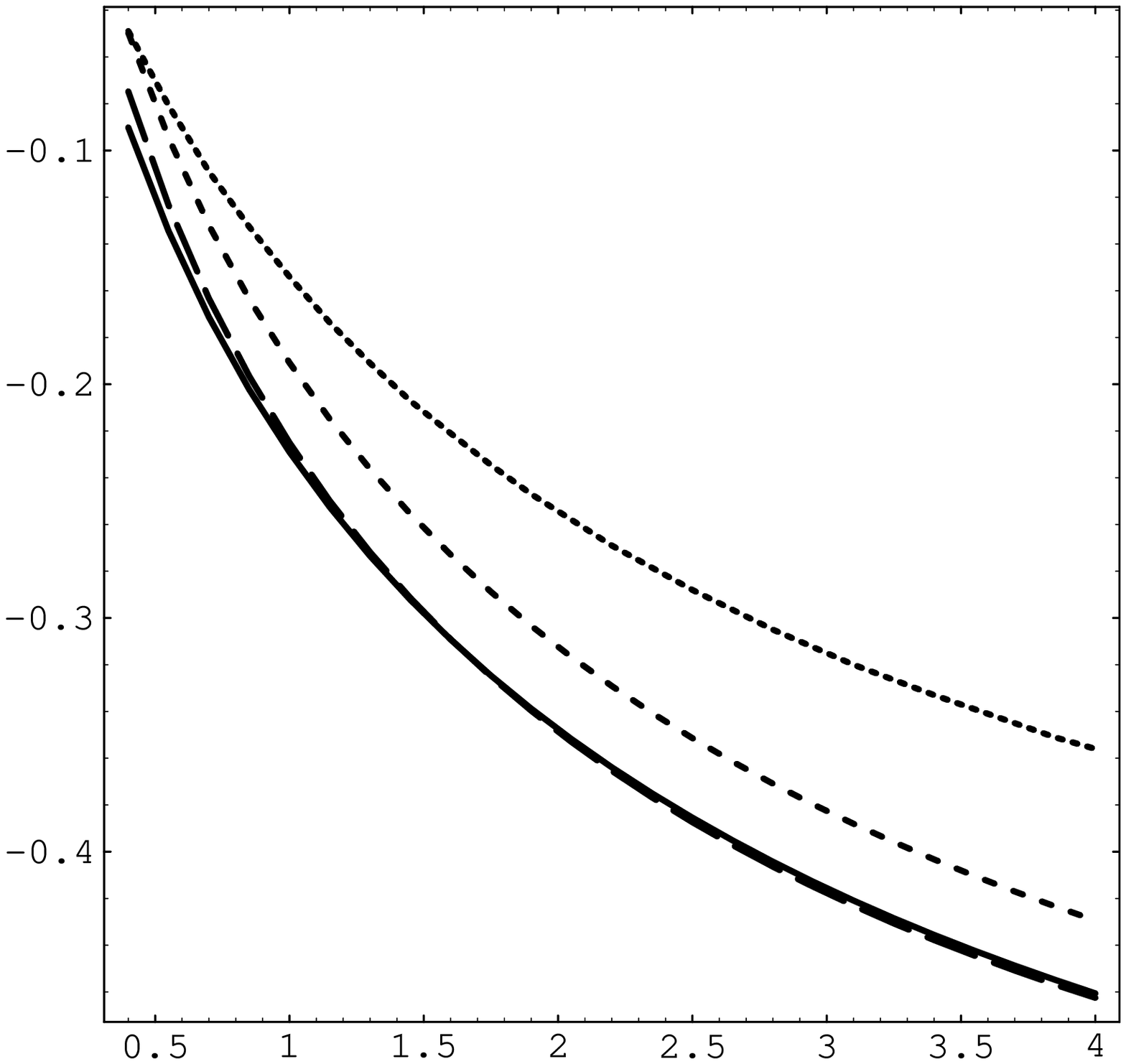,height=9.cm}$$

\vspace{-5.8cm}\null

\hspace*{3.6cm}
 \begin{sideways}
 {\small $ R_0 $ }
 \end{sideways}

\vspace{2.8cm}\null

\centerline{\small $\sqrt{s}\ (TeV)$}
\vspace{0.1cm}\null

\centerline{\small (c)}
\vspace{0.2cm}\null


\caption[7]{The ratio $R_0$ for
$\gamma\gamma\to b\bar b$ versus the energy; SM (a), 
MSSM($\tan\beta=4$) (b); MSSM($\tan\beta=40$) (c);
all logarithmic terms (solid),
leading terms only (small dashed), 
leading angular independent terms only(large dashed); 
all logarithmic without Yukawa terms (very small dashed).}
\label{R0b}
\end{figure}


\begin{figure}
$$\epsfig{file=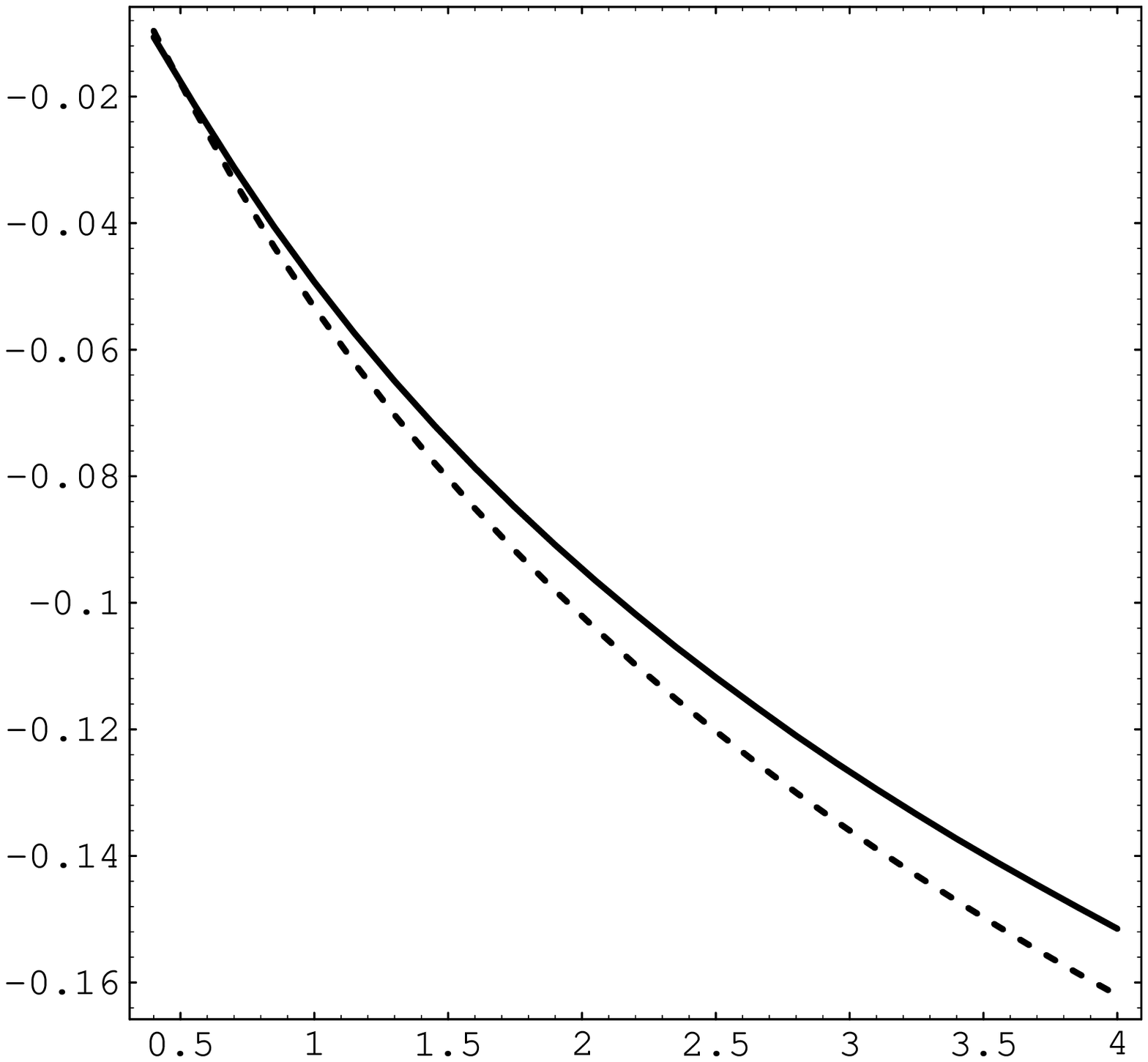,height=9.cm}\hspace{3.5cm}
\epsfig{file=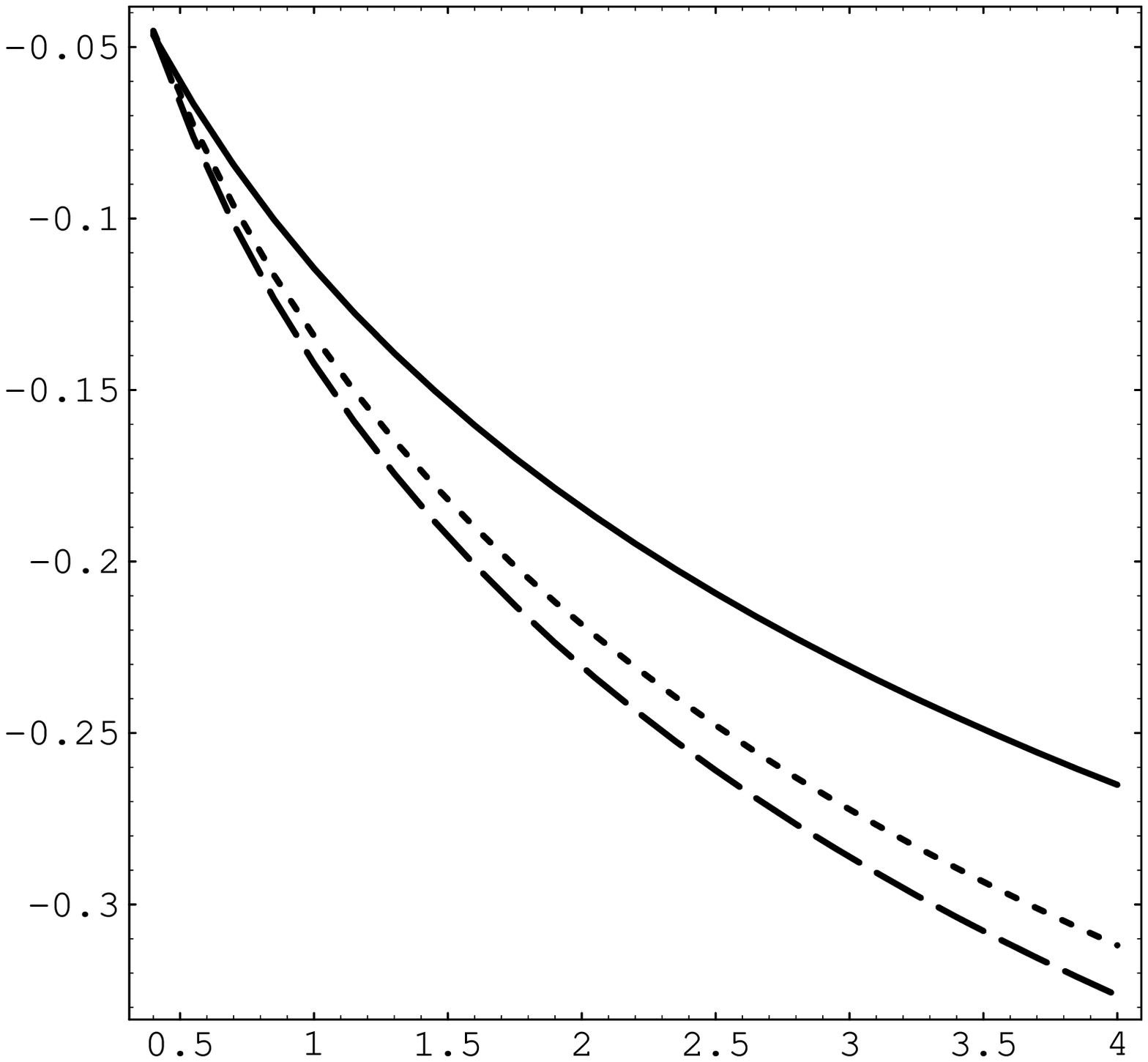,height=9.cm}$$
\vspace{-5.8cm}\null

\hspace*{-1.5cm}
 \begin{sideways}
 {\small $ R_0 $ }
 \end{sideways}
\hspace{9.5cm}
 \begin{sideways}
 {\small $ R_0 $ }
 \end{sideways}
\vspace{3.cm}\null

\centerline{\small $\sqrt{s}\ (TeV)$ \hspace{8.cm} \small $\sqrt{s}\ (TeV)$ }
\vspace{0.1cm}\null

\centerline{\small (a) \hspace{9.5cm} \small (b)}

$$\epsfig{file=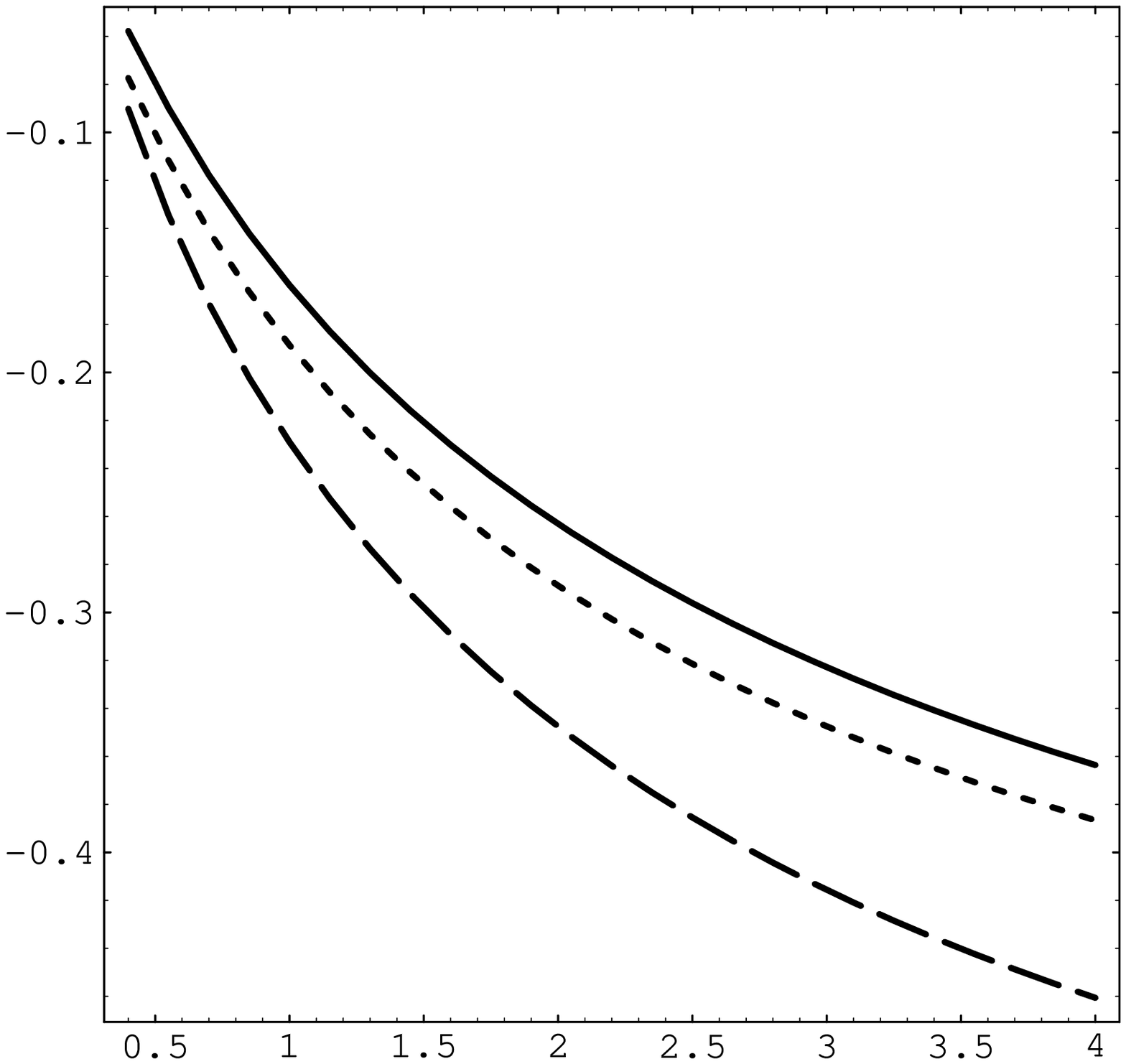,height=9.cm}$$

\vspace{-5.8cm}\null

\hspace*{3.6cm}
 \begin{sideways}
 {\small $ R_0 $ }
 \end{sideways}

\vspace{2.8cm}\null

\centerline{\small $\sqrt{s}\ (TeV)$}
\vspace{0.1cm}\null

\centerline{\small (c)}
\vspace{0.2cm}\null


\caption[8]{The ratio $R_0$ for
$\gamma\gamma\to f\bar f$ versus the energy; $l^+l^-$ (a),
$t\bar t$ (b), $b\bar b$ (c); SM (solid), MSSM ($\tan\beta=4$) (small
dashed), MSSM ($\tan\beta=40$) (large dashed).}
\label{R0}
\end{figure}


\begin{figure}[p]
$$\epsfig{file=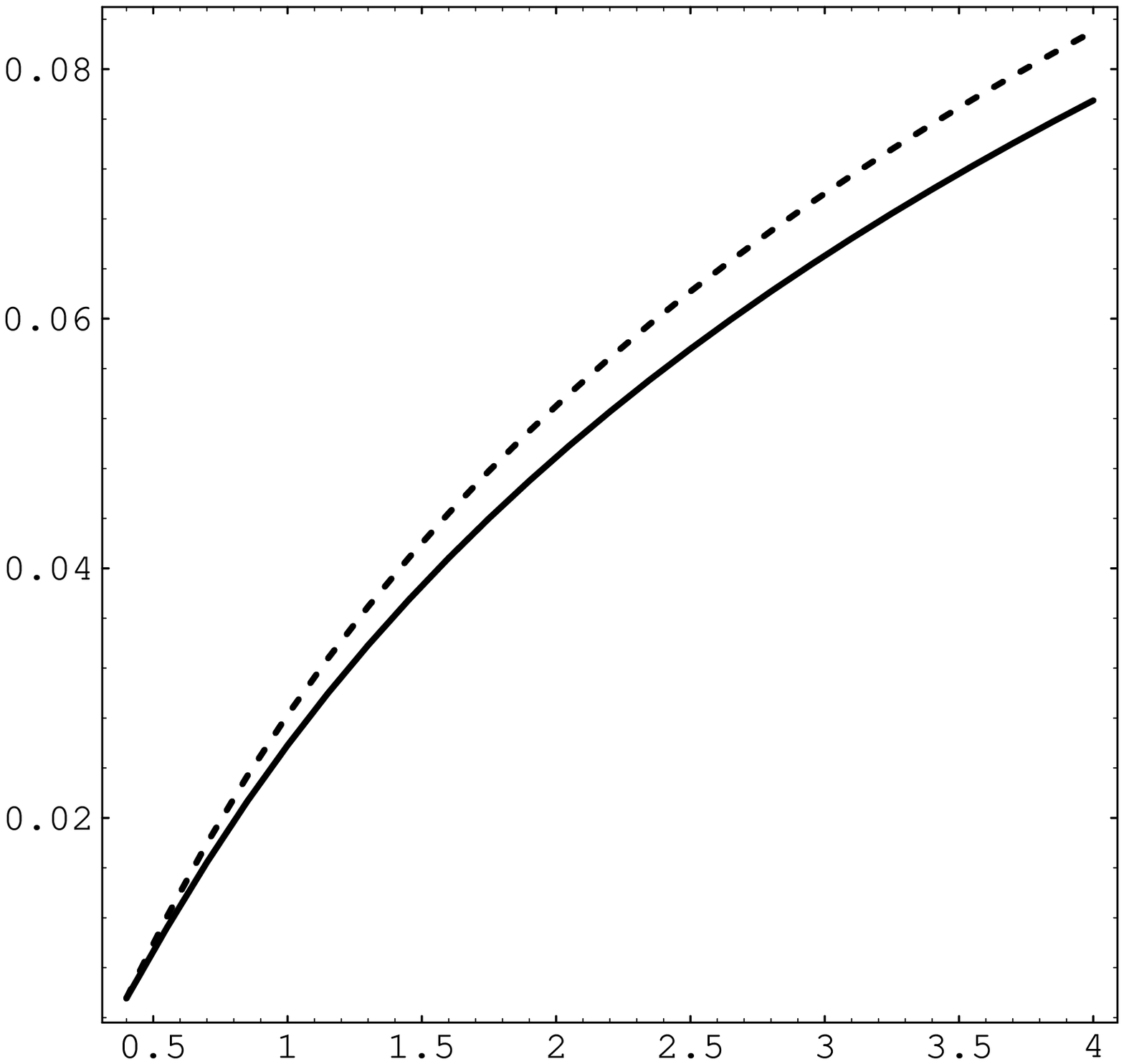,height=9.cm}\hspace{3.5cm}
\epsfig{file=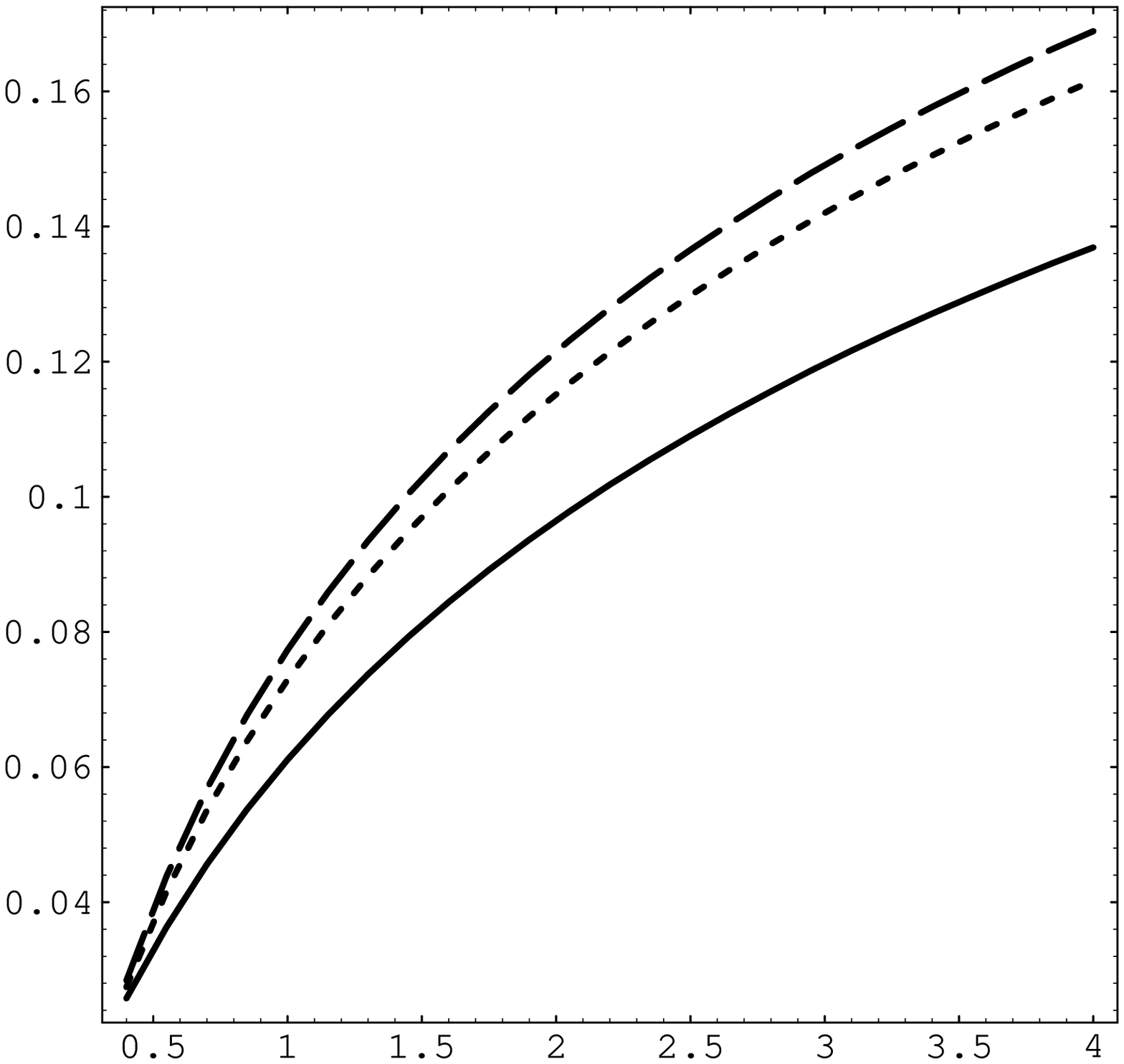,height=9.cm}$$
\vspace{-5.8cm}\null

\hspace*{-1.5cm}
 \begin{sideways}
 {\small $ R_{33} $ }
 \end{sideways}
\hspace{9.5cm}
 \begin{sideways}
 {\small $ R_{33} $ }
 \end{sideways}
\vspace{3.cm}\null

\centerline{\small $\sqrt{s}\ (TeV)$ \hspace{8.cm} \small $\sqrt{s}\ (TeV)$ }
\vspace{0.1cm}\null

\centerline{\small (a) \hspace{9.5cm} \small (b)}

$$\epsfig{file=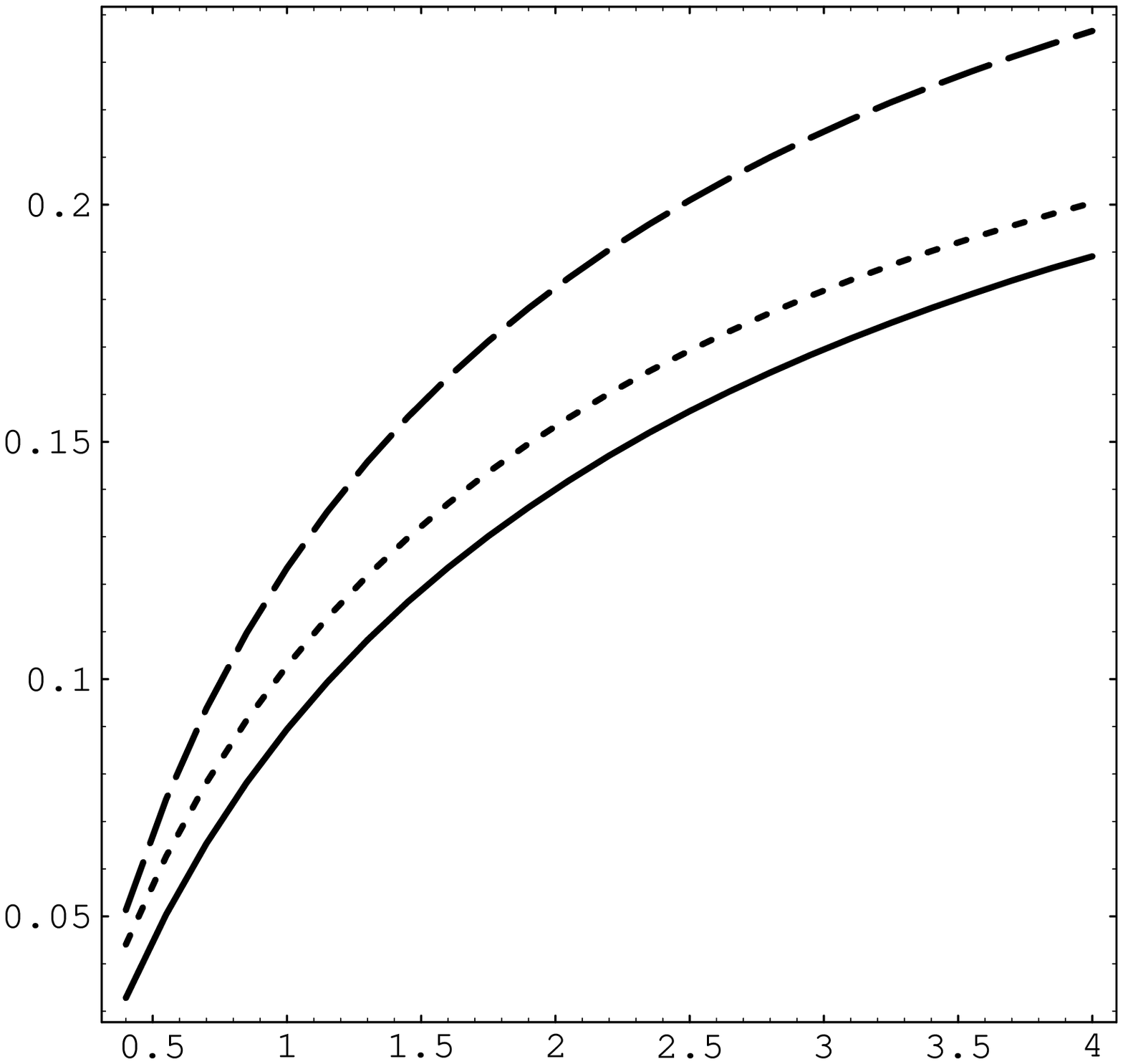,height=9.cm}$$

\vspace{-5.8cm}\null

\hspace*{3.6cm}
 \begin{sideways}
 {\small $ R_{33} $ }
 \end{sideways}

\vspace{2.8cm}\null

\centerline{\small $\sqrt{s}\ (TeV)$}
\vspace{0.1cm}\null

\centerline{\small (c)}
\vspace{0.2cm}\null


\caption[9]{The ratio $R_{33}$ for
$\gamma\gamma\to f\bar f$ versus the energy; $l^+l^-$ (a),
$t\bar t$ (b), $b\bar b$ (c); SM (solid), MSSM ($\tan\beta=4$) (small
dashed), MSSM ($\tan\beta=40$) (large dashed).}
\label{R33}
\end{figure}


\begin{figure}[p]
$$\epsfig{file=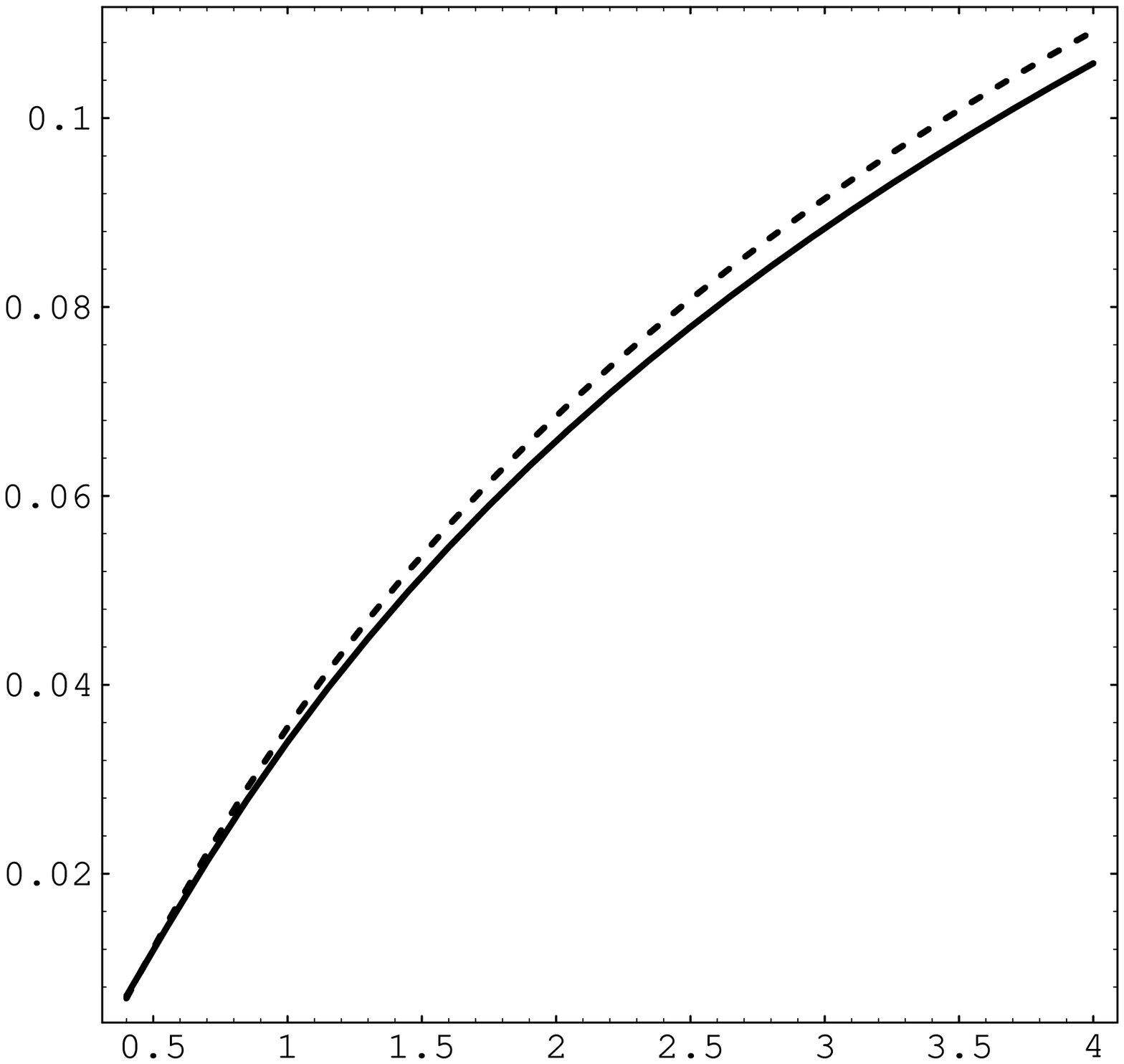,height=9.cm}\hspace{3.5cm}
\epsfig{file=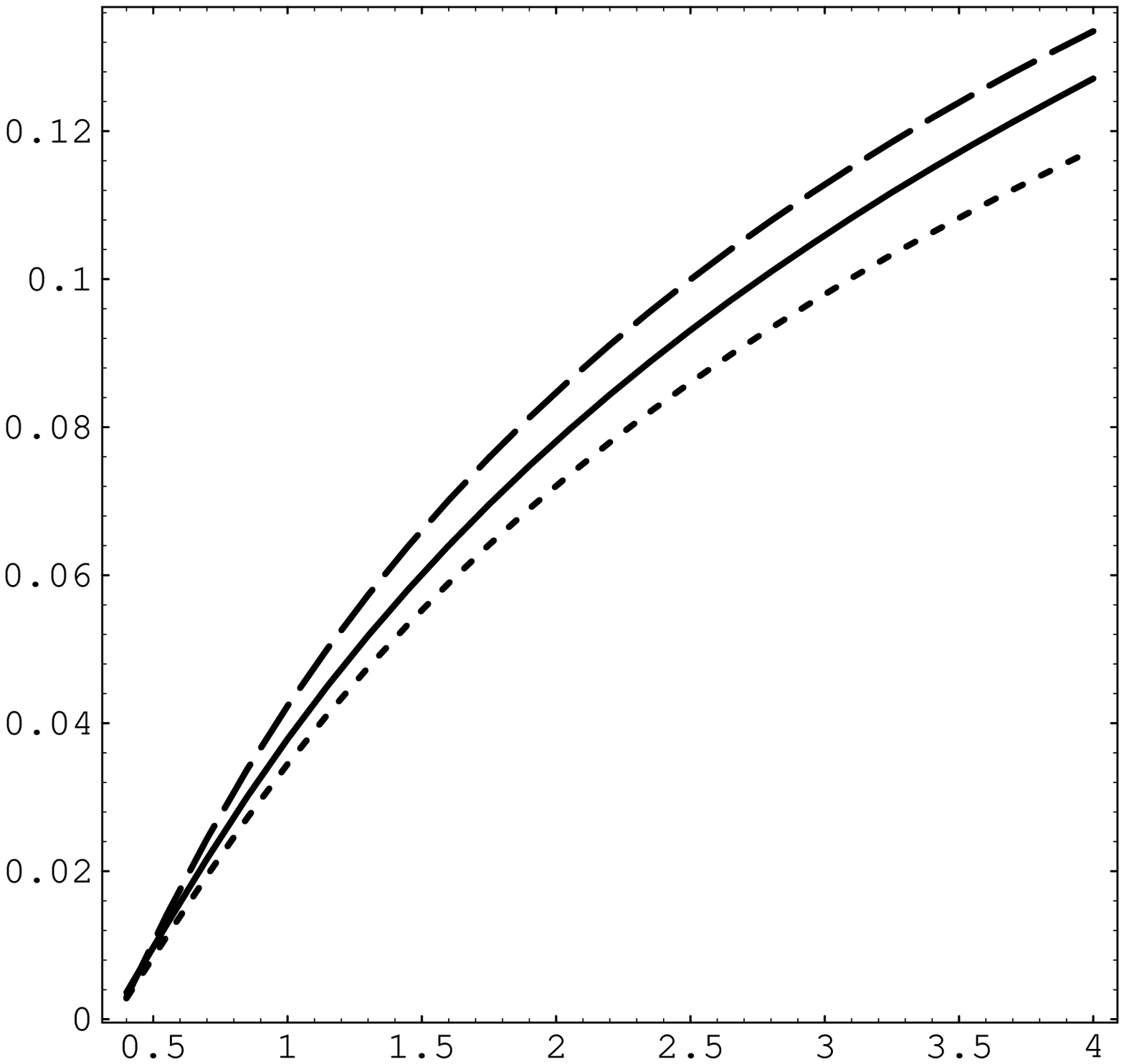,height=9.cm}$$
\vspace{-5.8cm}\null

\hspace*{-1.5cm}
 \begin{sideways}
 {\small $ R_{2} $ }
 \end{sideways}
\hspace{9.5cm}
 \begin{sideways}
 {\small $ R_{2} $ }
 \end{sideways}
\vspace{3.cm}\null

\centerline{\small $\sqrt{s}\ (TeV)$ \hspace{8.cm} \small $\sqrt{s}\ (TeV)$ }
\vspace{0.1cm}\null

\centerline{\small (a) \hspace{9.5cm} \small (b)}

$$\epsfig{file=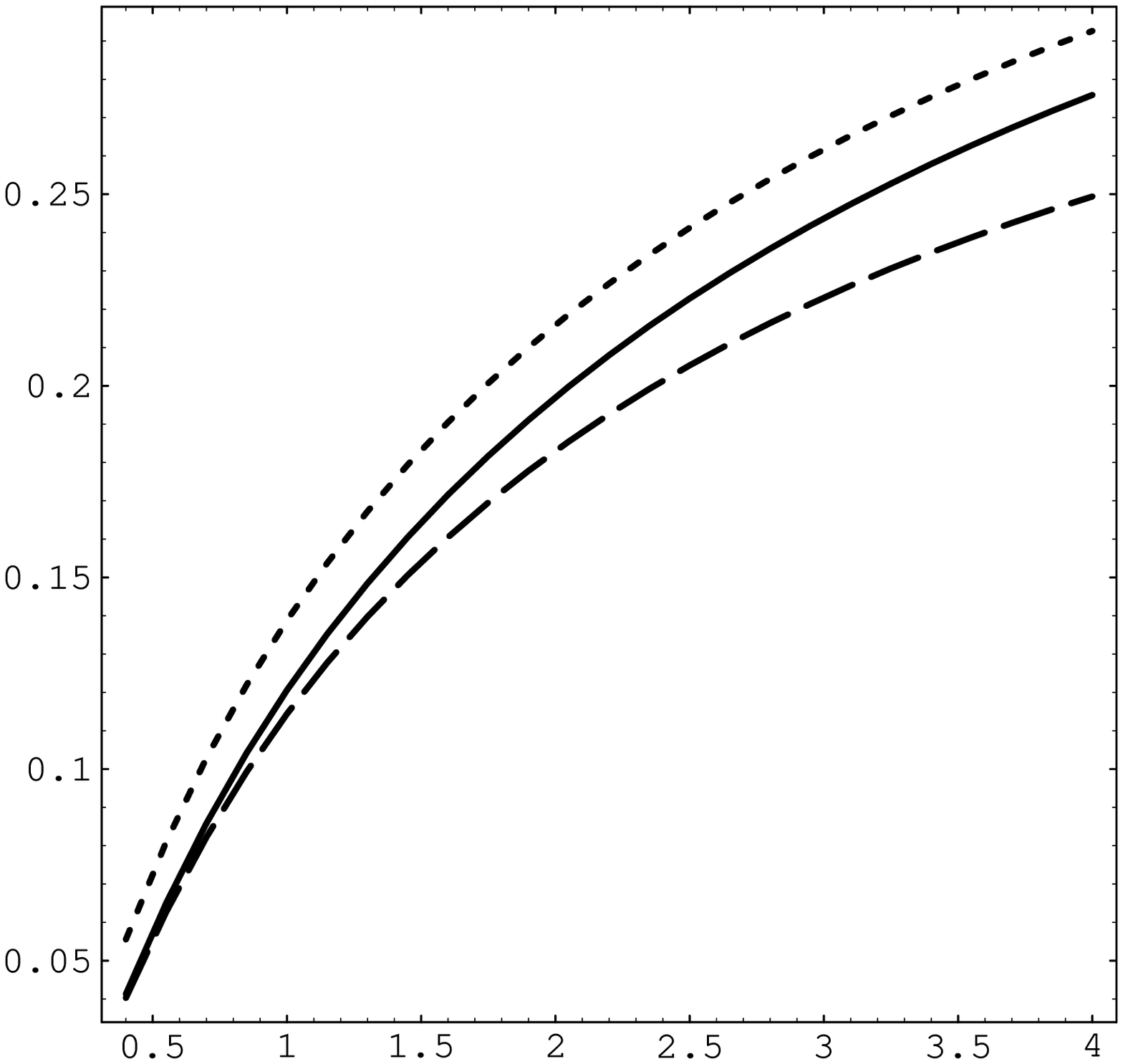,height=9.cm}$$

\vspace{-5.8cm}\null

\hspace*{3.6cm}
 \begin{sideways}
 {\small $ R_{2} $ }
 \end{sideways}

\vspace{2.8cm}\null

\centerline{\small $\sqrt{s}\ (TeV)$}
\vspace{0.1cm}\null

\centerline{\small (c)}
\vspace{0.2cm}\null


\caption[10]{The ratio $R_{2}$ for
$\gamma\gamma\to f\bar f$ versus the energy; $l^+l^-$ (a),
$t\bar t$ (b), $b\bar b$ (c); SM (solid), MSSM ($\tan\beta=4$) (small
dashed), MSSM ($\tan\beta=40$) (large dashed).}
\label{R2}
\end{figure}


\begin{figure}
$$\epsfig{file=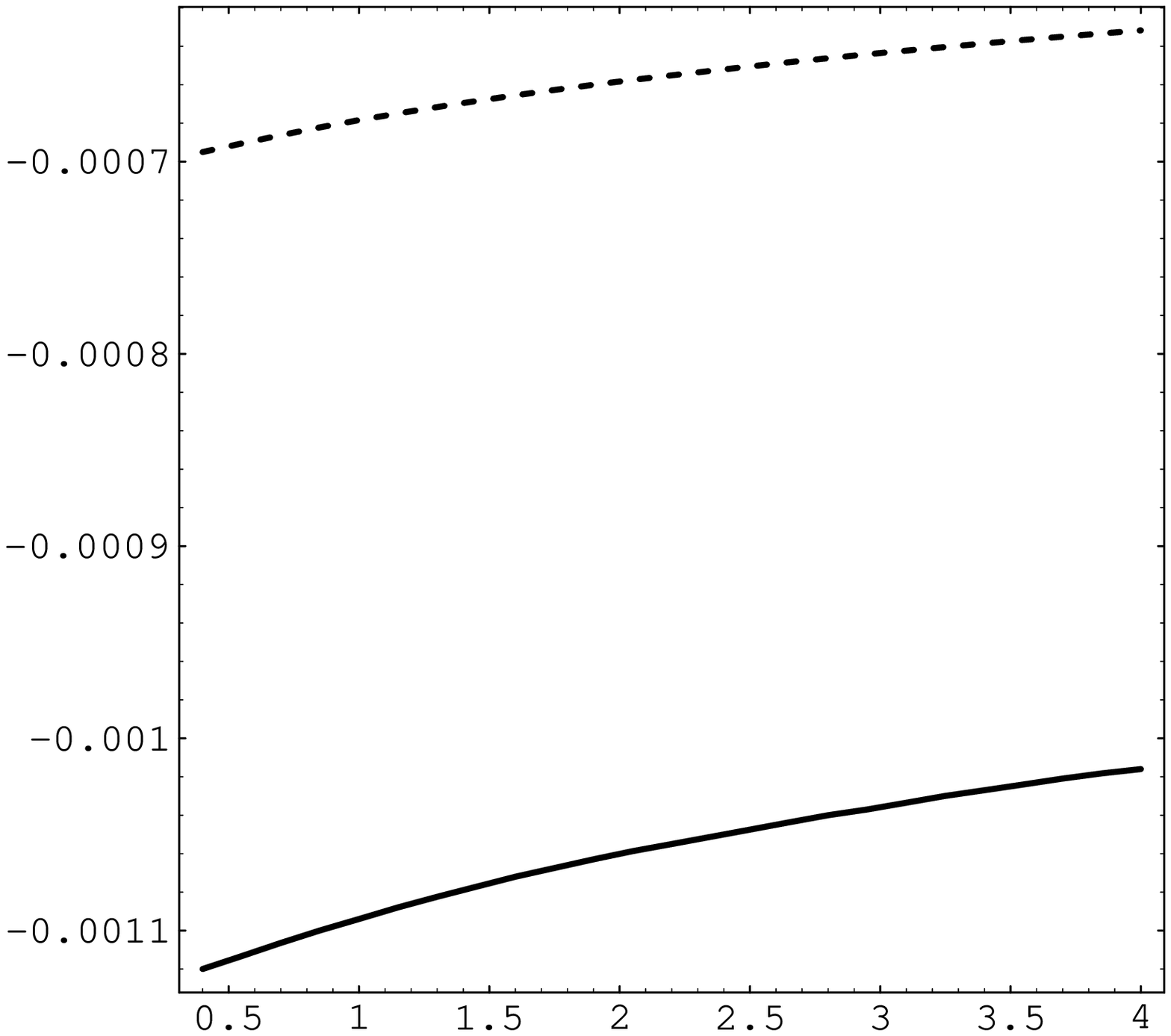,height=9.cm}\hspace{3.5cm}
\epsfig{file=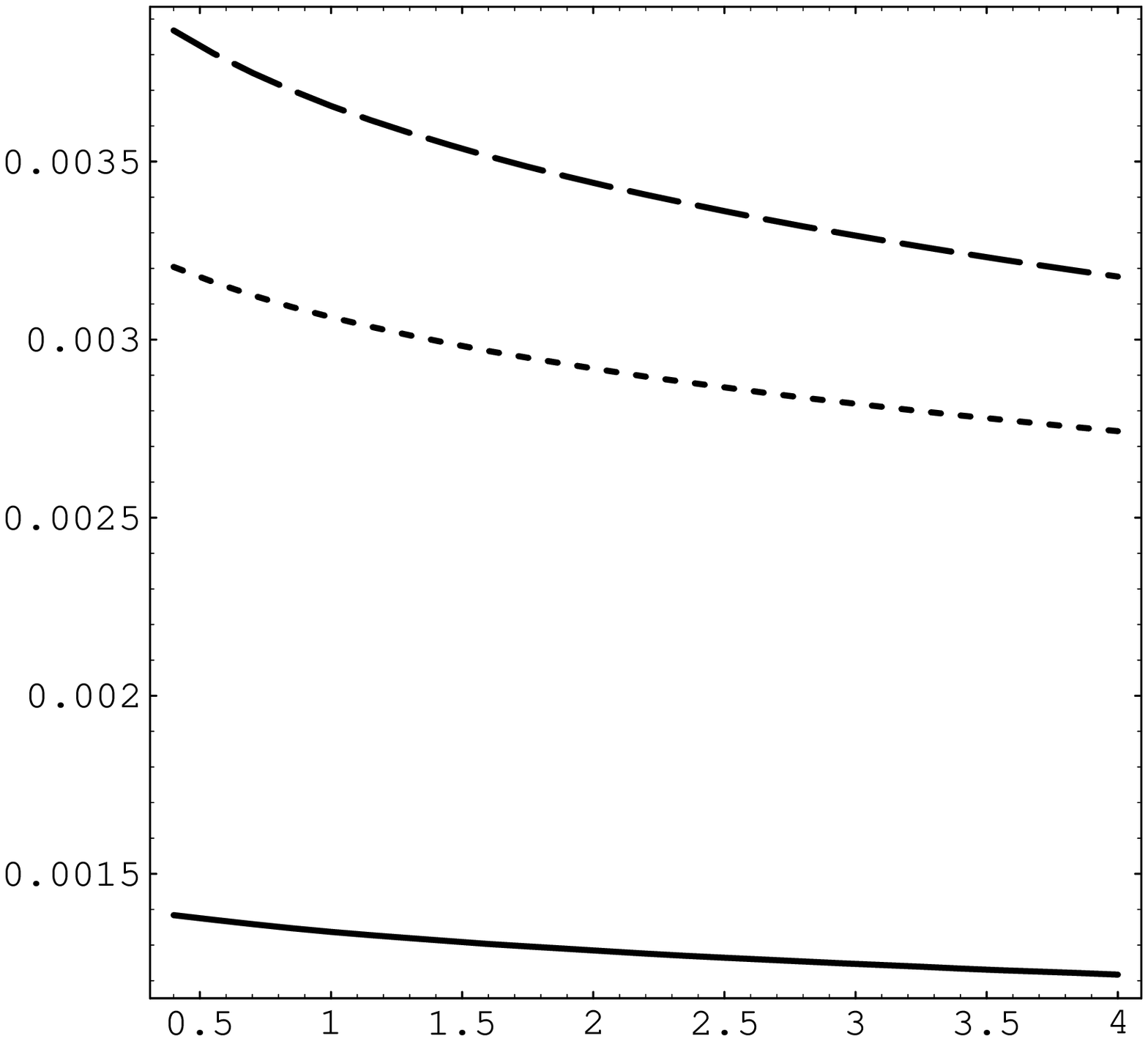,height=9.cm}$$
\vspace{-5.8cm}\null

\hspace*{-1.5cm}
 \begin{sideways}
 {\small $ R_{3} $ }
 \end{sideways}
\hspace{9.5cm}
 \begin{sideways}
 {\small $ R_{3} $ }
 \end{sideways}
\vspace{3.cm}\null

\centerline{\small $\sqrt{s}\ (TeV)$ \hspace{8.cm} \small $\sqrt{s}\ (TeV)$ }
\vspace{0.1cm}\null

\centerline{\small (a) \hspace{9.5cm} \small (b)}

$$\epsfig{file=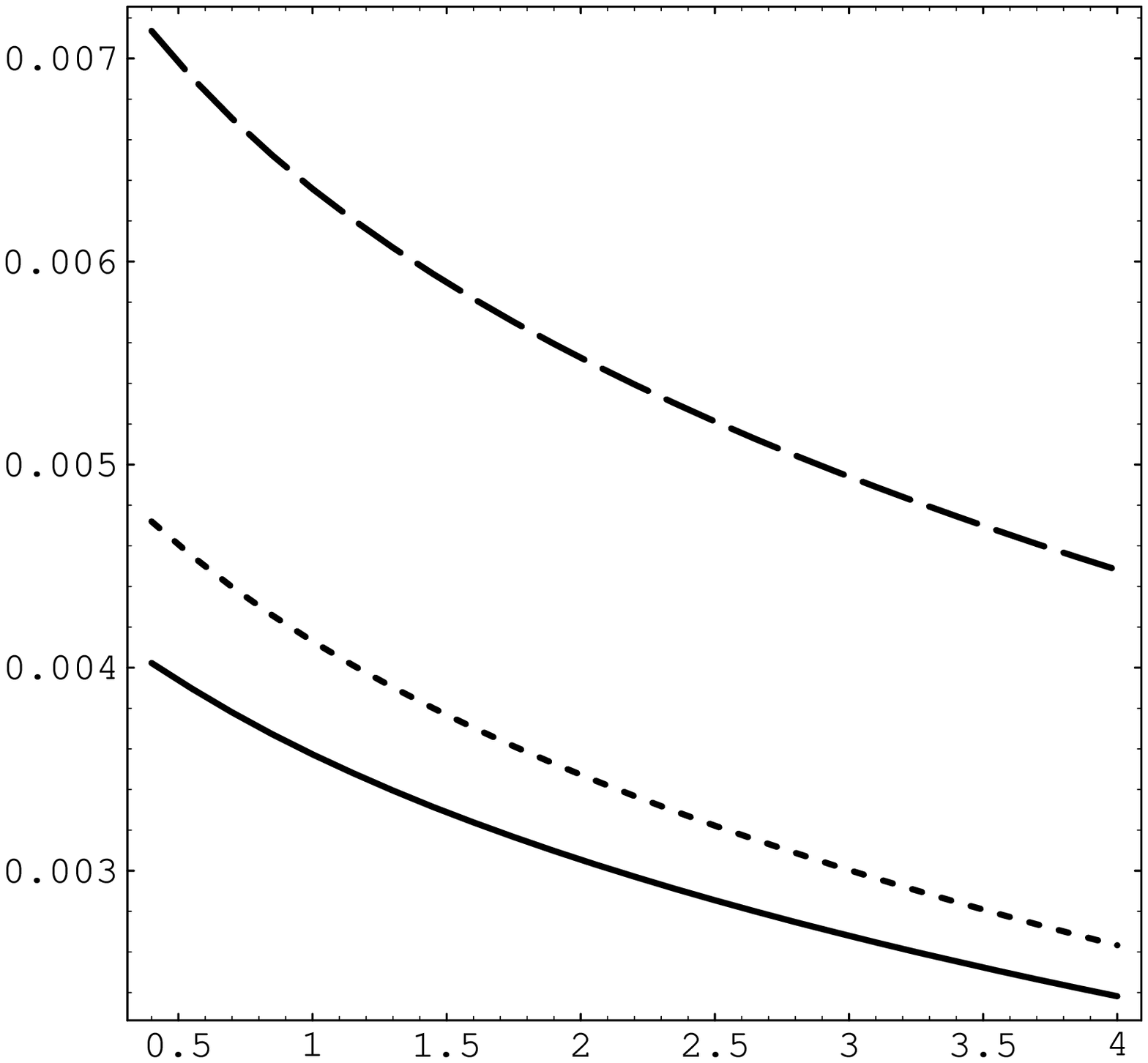,height=9.cm}$$

\vspace{-5.8cm}\null

\hspace*{3.6cm}
 \begin{sideways}
 {\small $ R_{3} $ }
 \end{sideways}

\vspace{2.8cm}\null

\centerline{\small $\sqrt{s}\ (TeV)$}
\vspace{0.1cm}\null

\centerline{\small (c)}
\vspace{0.2cm}\null


\caption[11]{The ratio $R_{3}$ for
$\gamma\gamma\to f\bar f$ versus the energy; $l^+l^-$ (a),
$t\bar t$ (b), $b\bar b$ (c); SM (solid), MSSM ($\tan\beta=4$) (small
dashed), MSSM ($\tan\beta=40$) (large dashed).}
\label{R3}
\end{figure}


\begin{figure}[p]
$$\epsfig{file=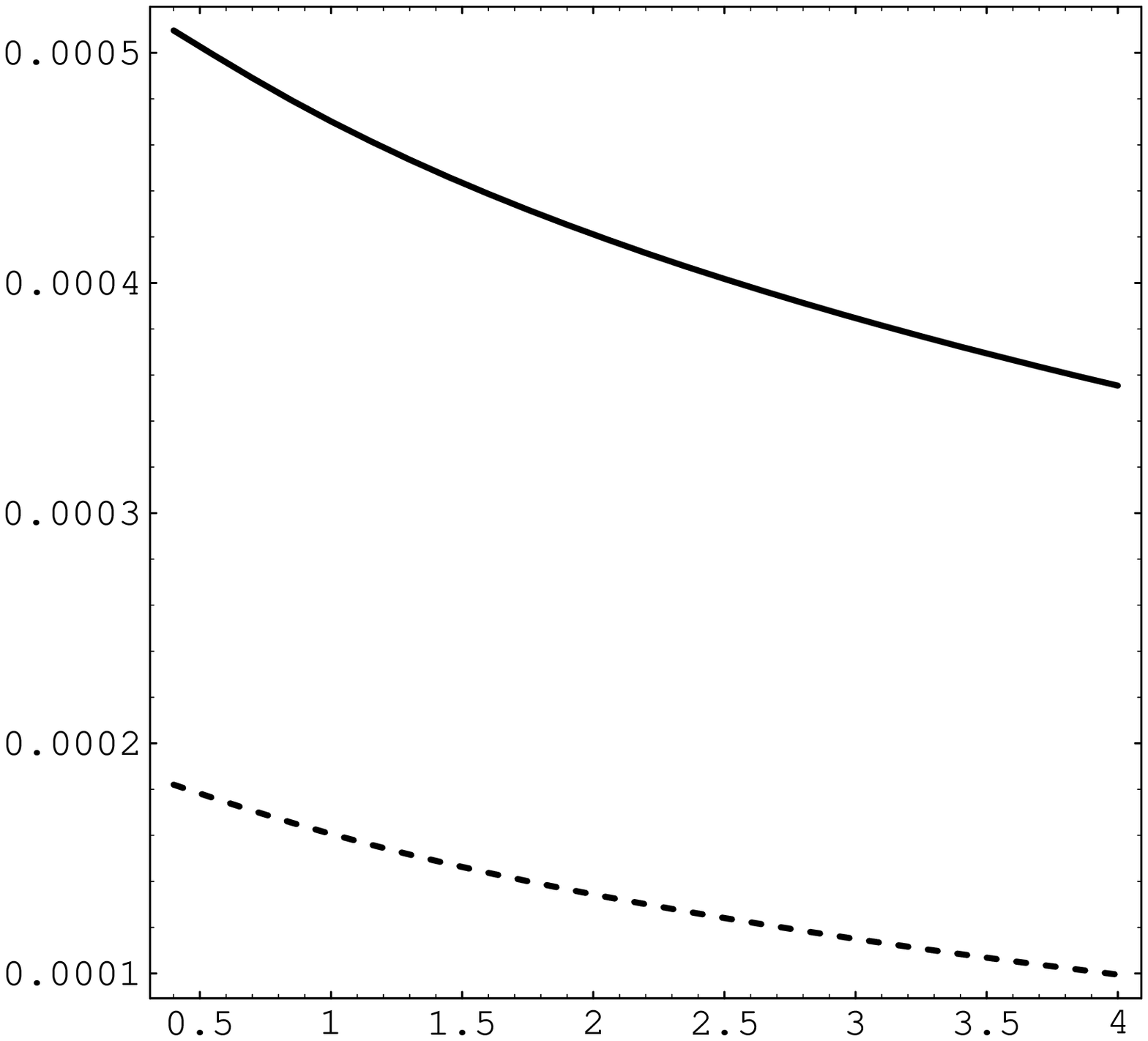,height=9.cm}\hspace{3.5cm}
\epsfig{file=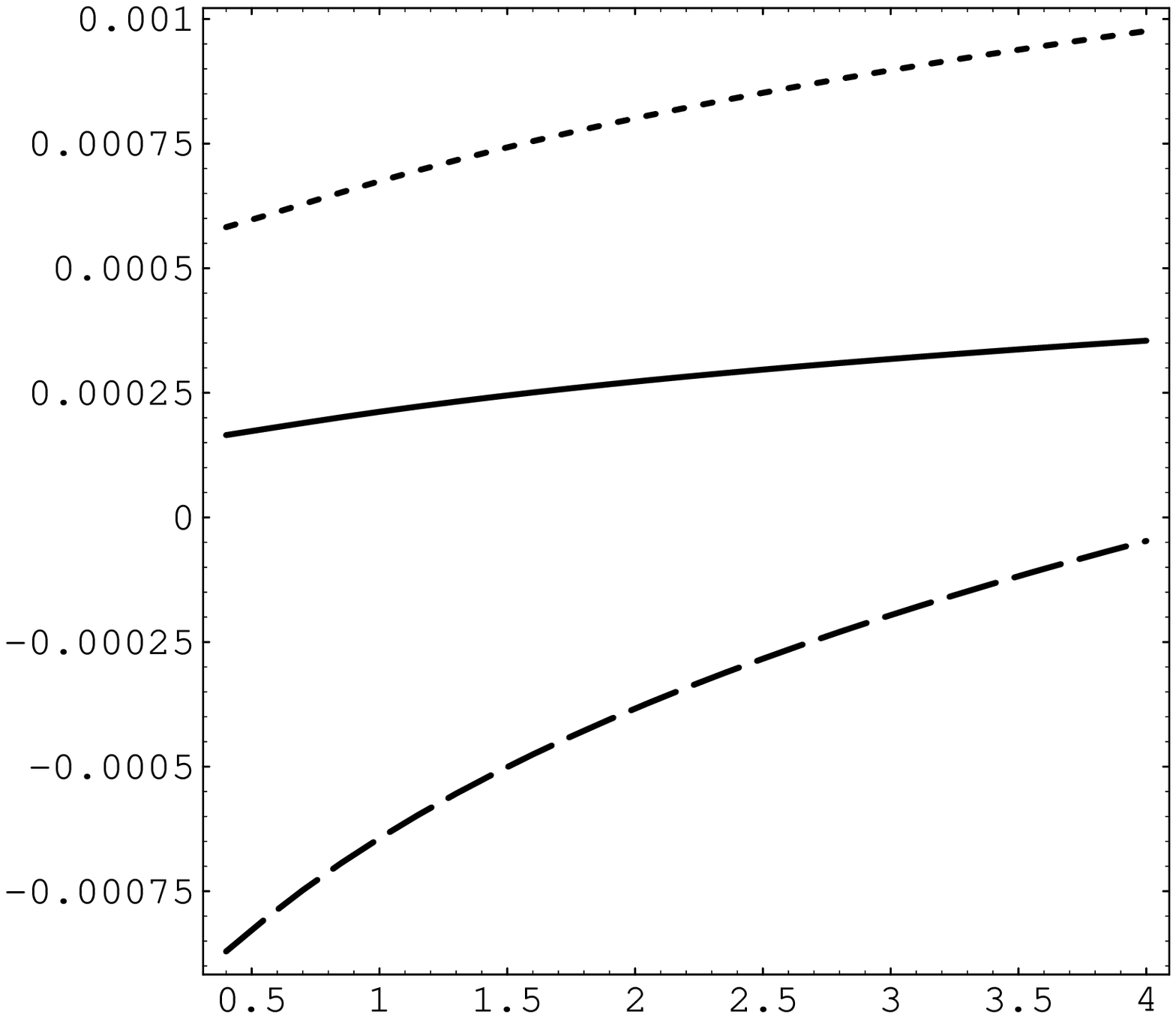,height=9.cm}$$
\vspace{-5.8cm}\null

\hspace*{-1.5cm}
 \begin{sideways}
 {\small $ R_{23} $ }
 \end{sideways}
\hspace{9.5cm}
 \begin{sideways}
 {\small $ R_{23} $ }
 \end{sideways}
\vspace{3.cm}\null

\centerline{\small $\sqrt{s}\ (TeV)$ \hspace{8.cm} \small $\sqrt{s}\ (TeV)$ }
\vspace{0.2cm}\null

\centerline{\small (a) \hspace{9.5cm} \small (b)}

$$\epsfig{file=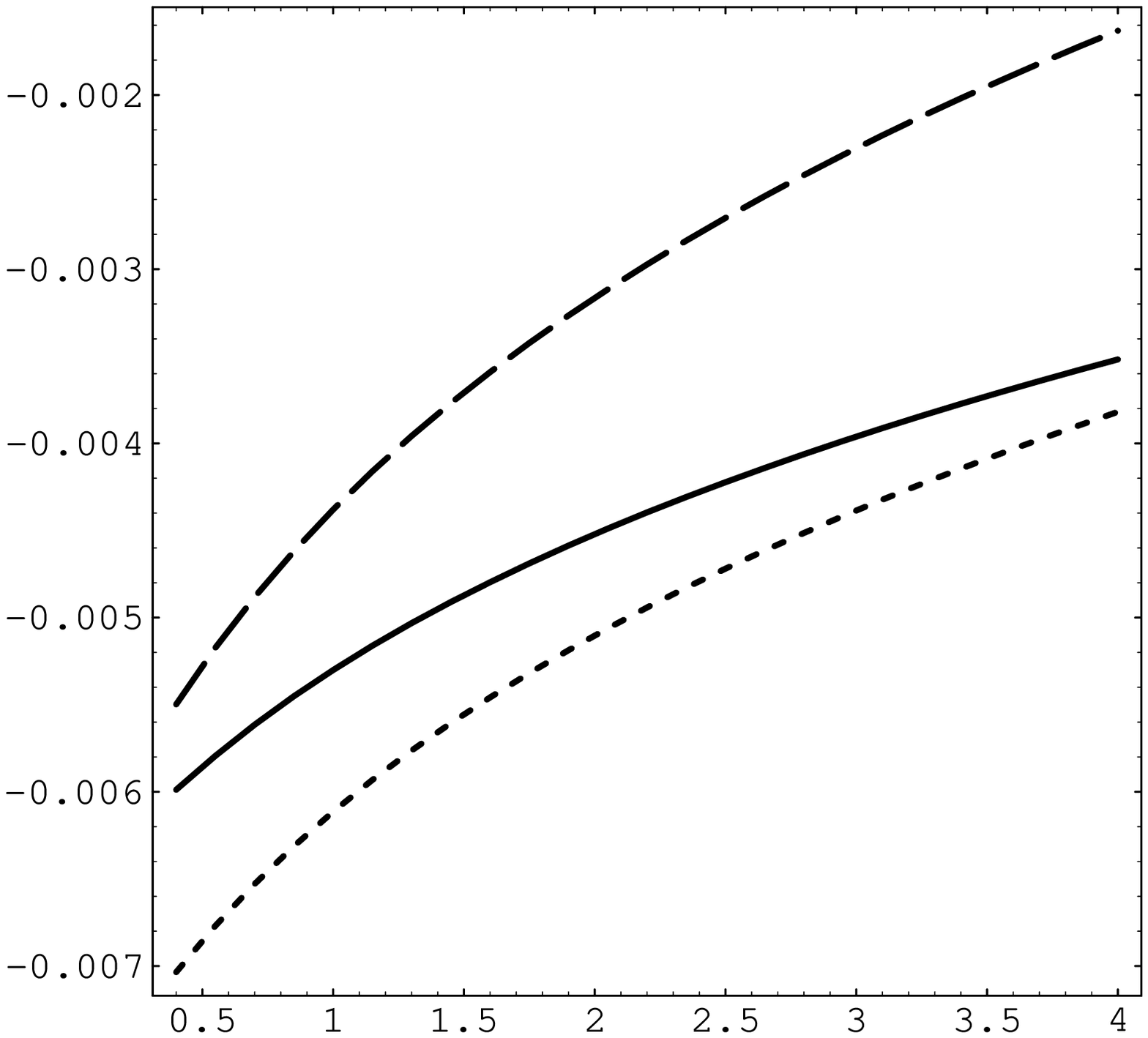,height=9.cm}$$

\vspace{-5.8cm}\null

\hspace*{3.6cm}
 \begin{sideways}
 {\small $ R_{23} $ }
 \end{sideways}

\vspace{2.8cm}\null

\centerline{\small $\sqrt{s}\ (TeV)$}
\vspace{0.1cm}\null

\centerline{\small (c)}
\vspace{0.2cm}\null


\caption[12]{The ratio $R_{23}$ for
$\gamma\gamma\to f\bar f$ versus the energy; $l^+l^-$ (a),
$t\bar t$ (b), $b\bar b$ (c); SM (solid), MSSM ($\tan\beta=4$) (small
dashed), MSSM ($\tan\beta=40$) (large dashed).}
\label{R23}
\end{figure}

\end{document}